%% file: paper.tex
\documentclass[sigconf,nonacm]{acmart}
\setcopyright{none}  

\acmConference{}{}{}
\acmBooktitle{}
\acmYear{} \acmDOI{} \acmISBN{}

\AtBeginDocument{%
  }

\usepackage{algorithm}
\usepackage{algpseudocode}
\usepackage{comment}
\usepackage{multirow}
\usepackage{booktabs}
\usepackage{amsmath}
\usepackage{float} 
\usepackage{graphicx}
\graphicspath{ {./images/} }
\usepackage{listings}
\usepackage{xcolor}
\usepackage{xspace}
\usepackage{multicol}
\usepackage{enumitem}
\usepackage{needspace}
\usepackage{subcaption}
\usepackage{tikz}
\usepackage{parskip}

\definecolor{codegray}{rgb}{0.5,0.5,0.5}
\definecolor{codepurple}{rgb}{0.58,0,0.82}
\definecolor{backcolour}{rgb}{0.95,0.95,0.92}

\lstdefinelanguage{SQL}{
  morekeywords={
    SELECT,FROM,WHERE,GROUP,BY,ORDER,HAVING,AS,ON,JOIN,INNER,LEFT,RIGHT,
    OUTER,UNION,ALL,INSERT,INTO,VALUES,UPDATE,SET,DELETE,CREATE,DROP,
    TABLE,VIEW,INDEX,AND,OR,NOT,NULL,IS,DISTINCT,CASE,WHEN,THEN,END,
    COUNT,PROMPT,FILE,TO_FILE,
    AI\_FILTER,AI\_AGG,AI\_CLASSIFY,AI\_COMPLETE,AI\_SUMMARIZE\_AGG
  },
  sensitive=false,
  morecomment=[l]{--},
  morestring=[b]',
}

\lstdefinestyle{sqlstyle}{
  commentstyle=\color{gray}\itshape,
  keywordstyle=\color{blue}\bfseries,
  stringstyle=\color{codepurple},
  basicstyle=\ttfamily\small,
  breaklines=true,
  frame=single,
  showstringspaces=false,
  captionpos=b,
  columns=flexible,
}

\newcommand{\sn}{SUPG\xspace}                    
\newcommand{\supgit}{SUPG-IT\xspace}             
\newcommand{\gamcal}{GAMCAL\xspace}              
\newcommand{\supgsp}{SUPG-SP\xspace}             

\newcommand{\supgitfull}{SUPG with Iterative Targeting (\supgit)}
\newcommand{\gamcalfull}{GAM-Calibrated Cascade (\gamcal)}

\newcommand{\SD}{\mathcal{D}}           
\newcommand{\SSe}{\mathcal{S}}          
\newcommand{\ex}{\mathbb{E}}            
\DeclareMathOperator*{\argmin}{arg\,min}
\DeclareMathOperator*{\argmax}{arg\,max}

\begin{document}
 
\title{Streaming Model Cascades for Semantic SQL}

\author{Paweł Liskowski}
\email{pawel.liskowski@snowflake.com}
\affiliation{%
  \institution{Snowflake Inc.}
  \city{Poznań}
  \country{Poland}
}

\author{Kyle Schmaus}
\email{kyle.schmaus@snowflake.com}
\affiliation{%
  \institution{Snowflake Inc.}
  \city{San Francisco, CA}
  \country{USA}
}

\renewcommand{\shortauthors}{Liskowski et al.}

\begin{abstract}
Modern data warehouses extend SQL with semantic operators that invoke large language models on each qualifying row, making per-row inference orders of magnitude more expensive than traditional SQL. Model cascades reduce this cost by routing most rows through a fast proxy model and delegating uncertain cases to an expensive oracle. Prior \sn-style cascades, however, require a global proxy-score pass that is itself an LLM-inference workload and blocks output in pipelined query engines. They also target either precision or recall and cannot serve workloads that need both. We formalize the cascade routing problem for streaming semantic SQL with independent parallel workers and present two complementary algorithms within this model. \supgit extends \sn from single-pass, single-metric estimation to streaming execution by iteratively refining two thresholds as oracle labels accumulate across batches, and is the first streaming cascade with joint probabilistic guarantees on user-specified precision and recall at a chosen failure probability $\delta$. \gamcal replaces user-specified targets with a single tradeoff parameter $\alpha$ between classification error and oracle cost, and learns a monotone Generalized Additive Model that calibrates proxy scores to true-positive probabilities and supplies pointwise uncertainty for stochastic routing. On six classification, filtering, and join benchmarks evaluated in a production semantic SQL engine, both algorithms reach $F_1 \geq 0.95$ at their best operating points. \gamcal also leads all six datasets at a 20\% delegation budget and reaches $F_1 \geq 0.95$ with up to 58\% fewer oracle calls than LOTUS's SUPG cascade. \supgit attains the highest best-case $F_1$, with a mean of 0.989 across the six datasets.
\end{abstract}

\maketitle

\section{Introduction}\label{sec:introduction}

Modern data warehouses increasingly integrate large language models (LLMs) directly into SQL through \emph{semantic operators} for filtering, classification, joins, and other relational operations driven by LLM-based reasoning~\cite{patel2025semantic, liskowski2025aisql, liu2025palimpzest}. These operators enable users to write declarative queries that blend relational operations with semantic reasoning: filtering customer reviews by sentiment, classifying documents into categories, or joining tables based on semantic similarity. However, each semantic operator invokes an LLM on every qualifying row: a single semantic filter applied to a million-row table may trigger hundreds of thousands of LLM calls, incurring costs that are orders of magnitude higher than traditional SQL operations and latencies measured in hours rather than seconds.

\emph{Model cascades} address this challenge by routing most rows through a fast, inexpensive \emph{proxy model} (e.g., a small LLM or embedding-based classifier) while escalating only uncertain cases to a powerful but expensive \emph{oracle model} (e.g., a large LLM). The proxy model produces a score for each row, and learned thresholds partition rows into regions that can be accepted, rejected, or delegated to the oracle. Cascades have demonstrated substantial cost reductions across LLM workloads (up to 98\% in API-based settings~\cite{chen2023frugalgpt} and up to 90\% in streaming inference~\cite{nie2024online}) and are recognized as a key optimization in semantic query processing systems~\cite{patel2025semantic, liu2025palimpzest}.

\sn~\cite{kang2020approximate} provides a statistical framework for learning such a threshold so that a user-specified recall or precision target is met with high probability. However, deploying \sn in production database systems reveals fundamental limitations along two axes. The first is \emph{architectural}: \sn assumes access to the entire dataset for computing normalized importance sampling weights, requiring a global pass over all proxy scores before any sampling can begin. On production inputs to a semantic filtering operator (\texttt{AI\_FILTER}), this global pass is itself an LLM-inference workload of millions of proxy calls that must complete before the first oracle delegation, blocking the cascade's output for the duration. Simple workarounds reproduce this barrier per worker (each worker still needs its complete proxy distribution before sampling can begin) or restore it globally through coordinated sampling, which requires re-deriving \sn's confidence bounds for the new sampling distribution. Furthermore, \sn estimates thresholds in a single pass with no mechanism to refine them as oracle labels accumulate across batches. Production semantic SQL engines instead require algorithms that operate in \emph{online, streaming mode}, refining estimates as batches arrive. A one-shot algorithm cannot meet this requirement: it either reintroduces the global barrier or estimates from each batch in isolation.

The second axis is \emph{methodological}. For binary predicates, \sn optimizes for a single metric (either recall or precision) but not both simultaneously. Realistic semantic-SQL workloads require both: an \texttt{AI\_FILTER} with 95\% recall but only 30\% precision delivers false positives to downstream \texttt{JOIN}s and aggregates, where they appear as spurious matches and biased counts. Without joint targeting, cascades cannot serve such workloads, which limits where semantic SQL itself can be deployed. The algorithm further assumes that proxy scores are well-calibrated, approximating the true probability $P(\text{positive} \mid x)$. In practice, proxy models are often poorly calibrated~\cite{guo2017calibration}, undermining the variance-optimality of \sn's importance weights and producing conservative thresholds with high delegation rates (Section~\ref{sec:gamcal-motivation}). Finally, users must specify fixed precision/recall targets that are often arbitrary, since different datasets have different inherent difficulty and users may not know \emph{a priori} what targets are achievable for their workload.

We present two algorithms that together address these architectural and methodological limitations:

\begin{enumerate}
    \item \textbf{\supgitfull} extends \sn to the streaming setting, refining two thresholds iteratively as oracle samples accumulate across batches while jointly targeting both precision and recall. The algorithm clips corrected targets to prevent over-delegation and collapses to a balanced threshold when precision-recall constraints conflict. Each worker processes its partition independently without inter-worker communication.
    
    \item \textbf{\gamcalfull} addresses the methodological limitations that \supgit inherits from the \sn framework: reliance on proxy calibration and the need for users to specify fixed accuracy targets. \gamcal \emph{learns} a calibration function mapping raw proxy scores to true probabilities using Generalized Additive Models (GAMs), then directly optimizes a cost-quality tradeoff controlled by a single parameter:
    \begin{equation}
        \min_{\tau_{\text{low}}, \tau_{\text{high}}} \; \alpha \cdot \text{error}(\tau_{\text{low}}, \tau_{\text{high}}) + (1 - \alpha) \cdot \text{cost}(\tau_{\text{low}}, \tau_{\text{high}})
    \end{equation}
    where $\tau_{\text{low}}$ and $\tau_{\text{high}}$ are the delegation thresholds and $\alpha \in [0, 1]$ is a weighting parameter. When $\alpha$ is high, the cascade prioritizes classification quality; when low, it minimizes oracle calls. The formulation adapts automatically to dataset difficulty, eliminating arbitrary target specification.
\end{enumerate}

The two algorithms serve complementary roles, each addressing needs the other cannot: \supgit is appropriate when users have explicit quality targets and want probabilistic guarantees, while \gamcal is appropriate when users prefer the system to automatically balance quality against cost without specifying any targets.

Both algorithms are designed for and evaluated in the context of Snowflake's Cortex AISQL, a production SQL engine that processes semantic operators over millions of rows~\cite{liskowski2025aisql}. On six benchmarks, both algorithms exceed $F_1 = 0.95$ at their best operating points. \gamcal outperforms the SUPG cascade in LOTUS~\cite{patel2025semantic} at cost-sensitive operating points, leading on all six datasets under a $20\%$ delegation budget and requiring up to $58\%$ fewer oracle calls to reach $F_1 \geq 0.95$. \supgit reaches the highest quality ceiling, with a mean peak $F_1$ of $0.989$.

In summary, our contributions are:
\begin{itemize}[leftmargin=*,itemsep=2pt]
    \item A formalization of the model cascade problem for streaming semantic SQL with independent parallel workers, introducing two complementary problem formulations (Section~\ref{sec:problem}).
    \item \supgit, the first cascade algorithm that provides joint precision-recall guarantees in online, streaming execution (Section~\ref{sec:supgit}).
    \item \gamcal, a calibration-based cascade that replaces user-specified targets with learned cost-quality optimization, using GAMs for probability calibration with uncertainty quantification (Section~\ref{sec:gamcal}).
\end{itemize}

The remainder of this paper is organized as follows. Section~\ref{sec:related} reviews related work on approximate selection, model cascades, and probability calibration. Section~\ref{sec:problem} formalizes the cascade routing problem. Sections~\ref{sec:supgit} and~\ref{sec:gamcal} present our two algorithms in detail. Section~\ref{sec:experiments} evaluates both algorithms on six real-world benchmarks. Section~\ref{sec:conclusion} concludes with limitations and future directions.

\section{Related Work}\label{sec:related}

\paragraph{Approximate selection with proxy models.}
Kang et al.~\cite{kang2020approximate} introduced SUPG, a framework for approximate data selection that uses importance-weighted oracle sampling to find a proxy score threshold meeting a user-specified recall or precision target with high probability. Earlier systems such as NoScope~\cite{kang2017noscope} and probabilistic predicates~\cite{lu2018accelerating} filtered data using proxy models but provided no statistical guarantees on result quality. SUPG addressed this gap through confidence bounds and variance-optimal importance sampling with weights proportional to the square root of proxy scores. Our work builds on SUPG while removing two assumptions: that the entire dataset is available before sampling begins, and that only a single metric is targeted. \gamcal further departs from the SUPG framework by replacing statistical threshold estimation with learned probability calibration.

BARGAIN~\cite{zeighami2025bargain} improves \sn's statistical estimator along an axis orthogonal to ours. It replaces \sn's importance sampling with adaptive iteration over candidate thresholds and tightens \sn's CLT-based asymptotic guarantees into finite-sample betting bounds~\cite{waudby2024estimating}. The architectural critique we develop in Section~\ref{sec:introduction} applies to BARGAIN as well. The global-access barrier in particular applies more strongly: \sn requires a single global pass to normalize importance weights, whereas BARGAIN's multi-round adaptive sampling draws fresh oracle labels at each iteration based on observations from earlier rounds. \supgit's iterative refinement schedule applies to any per-batch confidence-bound estimator. Replacing \sn's CLT-based bounds with BARGAIN's tighter betting bounds inside this schedule is a natural extension that we leave to future work.

\paragraph{LLM cascades.}
Routing queries through progressively more capable models has been explored extensively. FrugalGPT~\cite{chen2023frugalgpt} learns a scoring function that predicts answer reliability and cascades queries through increasingly expensive API endpoints. Nie et al.~\cite{nie2024online} formalize cascade learning as online imitation learning where smaller models learn from LLM demonstrations in a streaming setting. Jitkrittum et al.~\cite{jitkrittum2023confidence} characterize when confidence-based deferral is optimal and identify failure modes involving specialist models and label noise. Wang et al.~\cite{wang2024cascadeaware} modify the small model's training loss to focus on tokens that at least one cascade model predicts correctly. Zellinger and Thomson~\cite{zellinger2025rational} model the joint distribution of calibrated confidences across a sequence of LLMs using Markov copulas for continuous threshold optimization.

Our setting differs from these methods in two respects. First, we operate on binary predicates embedded in SQL, where the proxy produces a continuous confidence score rather than a discrete prediction or generated text. Second, our algorithms must run in a streaming, per-partition execution model imposed by distributed database architectures, whereas most LLM cascade work assumes batch access to a validation set or centralized routing.

\paragraph{Probability calibration.}
Mapping model outputs to well-calibrated probabilities is a long-standing problem. Platt scaling~\cite{platt1999probabilistic} fits a logistic function to raw scores, temperature scaling~\cite{guo2017calibration} adjusts a single parameter, and isotonic regression~\cite{zadrozny2002transforming} fits a non-parametric monotone map. Guo et al.~\cite{guo2017calibration} demonstrated that modern neural networks are often poorly calibrated despite high accuracy. Spline-based methods generalize these approaches: Generalized Additive Models~\cite{hastie1986generalized} fit penalized smoothing splines, and Lucena~\cite{lucena2018splinecalib} and Wang and Dunson~\cite{wang2011ispline} have applied cubic and monotone-spline calibration respectively. Dormann~\cite{dormann2020calibration} surveys spline-based against parametric calibration methods. Calibration has also been applied within cascade frameworks: Zellinger and Thomson~\cite{zellinger2024hcma} combine a nonlinear log-transform with Platt scaling to improve routing decisions in LLM cascades with abstention. \gamcal uses calibration for a different purpose: it trains a GAM on oracle samples to predict expected quality metrics for any threshold pair in closed form, enabling direct threshold optimization without additional oracle queries.

\paragraph{Semantic query processing.}
Several systems embed LLM-powered operators into SQL and dataframe APIs for processing unstructured data at scale~\cite{patel2025semantic, liskowski2025aisql, liu2025palimpzest}. LOTUS~\cite{patel2025semantic} optimizes semantic filters with a \sn-based cascade that learns two proxy-score thresholds to meet joint precision and recall targets at a user-specified failure probability, delegating records in the intermediate uncertain region to the oracle. Because it materializes proxy scores over the entire table before sampling, the cascade inherits the global-access barrier described in Section~\ref{sec:introduction}. We compare against a faithful streaming reimplementation of the LOTUS cascade (\supgsp) in Section~\ref{sec:experiments}. An alternative line of work bypasses cascading entirely by training lightweight proxy models to replace LLM invocations: UQE~\cite{dai2024uqe} uses embedding-based classifiers for semantic filters, and Chung et al.~\cite{chung2026proxy} demonstrate order-of-magnitude cost reductions through full proxy replacement in BigQuery and AlloyDB. These approaches provide no statistical guarantees on precision or recall, relying instead on heuristic quality thresholds for fallback. Our cascades occupy a middle ground: they route individual records between proxy and oracle models with statistical quality guarantees (\supgit) or learned cost-quality optimization (\gamcal), accepting higher per-predicate cost than full replacement in exchange for controlled accuracy.

\section{Problem Formulation}\label{sec:problem}

We formalize the model cascade problem for binary semantic predicates, establishing a two-threshold decision framework, two complementary optimization objectives, and the streaming execution model required for deployment in distributed database systems.

\subsection{Setting and Notation}

Consider a dataset $\SD = \{x_1, x_2, \ldots, x_n\}$ of $n$ records to be processed by a binary semantic predicate (e.g., \texttt{AI\_FILTER}). We have access to two models:

\begin{itemize}
    \item \textbf{Oracle model} $O: \SD \rightarrow \{0, 1\}$: An expensive but accurate model (e.g., a large LLM) that produces a binary label $y_i = O(x_i)$ for each record, at per-invocation cost $c_O$.
    
    \item \textbf{Proxy model} $A: \SD \rightarrow [0, 1]$: A fast, inexpensive model that returns a confidence score $A(x_i) \in [0, 1]$ estimating the probability that $y_i = 1$, at per-invocation cost $c_A \ll c_O$.
\end{itemize}

Following standard practice in the cascade literature~\cite{kang2020approximate}, we treat oracle labels as ground truth and measure all quality metrics with respect to $\{y_i\}$. The assumption is reasonable when the oracle is substantially more accurate than the proxy, as is typically the case for large versus small language models on semantic predicates~\cite{liskowski2025aisql}. Our goal is to produce predictions $\hat{y}_i$ for all records while minimizing oracle invocations. Since $c_A \ll c_O$, we assume the proxy is executed on all records, yielding scores $\{A(x_i)\}_{i=1}^n$.

\subsection{Two-Threshold Decision Framework}

Our cascade algorithms partition records into three regions using two thresholds $\tau_{\text{low}}$ and $\tau_{\text{high}}$ where $0 \leq \tau_{\text{low}} \leq \tau_{\text{high}} \leq 1$:

\begin{enumerate}
    \item \textbf{Reject region} ($A(x) < \tau_{\text{low}}$): The proxy is confident the record does not satisfy the predicate. Predict $\hat{y} = 0$ without oracle evaluation.
    
    \item \textbf{Accept region} ($A(x) \geq \tau_{\text{high}}$): The proxy is confident the record satisfies the predicate. Predict $\hat{y} = 1$ without oracle evaluation.
    
    \item \textbf{Uncertain region} ($\tau_{\text{low}} \leq A(x) < \tau_{\text{high}}$): The proxy lacks confidence. Invoke the oracle to obtain $\hat{y} = O(x)$.
\end{enumerate}

The \emph{delegation rate} $d$ is the fraction of records routed to the oracle:
\begin{equation}
    d = \frac{|\{x_i : \tau_{\text{low}} \leq A(x_i) < \tau_{\text{high}}\}|}{n}
\end{equation}

We state the framework in terms of the raw proxy score $A(x)$ for concreteness. More generally, the thresholds may operate on any monotone transformation of the proxy output. \gamcal (Section~\ref{sec:gamcal}) replaces $A(x)$ with a calibrated decision score that incorporates learned probability estimates (Section~\ref{sec:gamcal-calibration}), but the three-region structure, quality metrics, and delegation rate carry over unchanged.

\subsection{Quality Metrics}

The cascade's three-region structure determines how classification errors arise. Accepted records receive $\hat{y}_i = 1$ without oracle verification; those with $y_i = 0$ are false positives. Rejected records receive $\hat{y}_i = 0$; those with $y_i = 1$ are false negatives. Delegated records receive $\hat{y}_i = y_i$ from the oracle and contribute no errors.

Errors thus originate exclusively from the accept and reject regions. The two thresholds play asymmetric, complementary roles: $\tau_{\text{high}}$ governs precision (raising it reduces false positives but delegates more records to the oracle), while $\tau_{\text{low}}$ governs recall (lowering it reduces false negatives at the cost of higher delegation). We quantify quality using:
\begin{align}
    \text{Precision} &= \frac{\text{TP}}{\text{TP} + \text{FP}}, \qquad
    \text{Recall} = \frac{\text{TP}}{\text{TP} + \text{FN}} \\
    F_\beta &= (1 + \beta^2) \cdot \frac{\text{Precision} \cdot \text{Recall}}{\beta^2 \cdot \text{Precision} + \text{Recall}}
\end{align}
where $F_\beta$ balances precision and recall ($\beta > 1$ favors recall; $\beta = 1$ yields the standard $F_1$ score). The cascade problem is thus a three-way tradeoff among precision, recall, and oracle cost.

\subsection{Problem Formulations}

We consider two complementary problem formulations, each addressed by one of our algorithms.

\subsubsection{Target-Based Formulation (\supgit)}

Users specify minimum precision and recall targets $t_P$ and $t_R$ along with a failure probability $\delta$. The goal is to find thresholds $(\tau_{\text{low}}, \tau_{\text{high}})$ such that, over the randomness of oracle sampling:
\begin{align}
    \Pr[\text{Precision} \geq t_P] &\geq 1 - \delta \\
    \Pr[\text{Recall} \geq t_R] &\geq 1 - \delta
\end{align}
while minimizing the delegation rate. The two constraints are coupled: improving recall requires lowering $\tau_{\text{low}}$ while improving precision requires raising $\tau_{\text{high}}$, and both adjustments widen the uncertain region. Satisfying both constraints simultaneously is harder than optimizing either in isolation, particularly when the proxy score distributions of positive and negative records overlap.

\subsubsection{Cost-Quality Tradeoff Formulation (\gamcal)}

Rather than specifying fixed targets, users provide a tradeoff parameter $\alpha \in [0, 1]$ controlling the balance between classification quality and oracle cost. The objective is:
\begin{equation}
    \min_{\tau_{\text{low}}, \tau_{\text{high}}} \; \alpha \cdot \text{error}(\tau_{\text{low}}, \tau_{\text{high}}) + (1 - \alpha) \cdot \text{cost}(\tau_{\text{low}}, \tau_{\text{high}})
    \label{eq:gamcal-objective}
\end{equation}

where $\text{error}$ captures classification quality (derived from $1 - F_\beta$, normalized for comparability as detailed in Section~\ref{sec:gamcal-optimization}) and $\text{cost}$ is the delegation rate. The formulation adapts to dataset difficulty: easy datasets achieve low error with low cost, while difficult datasets require higher delegation rates.

\subsection{Streaming Execution Model}\label{sec:streaming}

Unlike batch algorithms that access the entire dataset simultaneously, our algorithms operate in a \emph{streaming} setting motivated by distributed database execution (Figure~\ref{fig:streaming-model}). Data is partitioned across $W$ parallel workers, worker $w$ processing its partition as a sequence of batches $B_1^{(w)}, B_2^{(w)}, \ldots$. The execution model imposes one fundamental constraint: \emph{batches are processed incrementally and cannot be revisited}. A worker's state after processing batch $t$ may depend on its previous state and the current batch, but not on future batches or previously processed ones. 

The constraint applies to \emph{raw data}: a worker may not re-read records from earlier batches. However, lightweight derived quantities (such as accumulated oracle samples and threshold estimates) are retained in the worker's state and remain available throughout execution.

\input{tikz-streaming-model}

The algorithms presented in this paper satisfy a stronger property: each worker operates \emph{independently}, without sharing samples, exchanging threshold estimates, or synchronizing with other workers. Independence eliminates inter-worker communication overhead and enables straightforward deployment in production SQL engines where partitions are processed in isolation. It also precludes reliance on global statistics (e.g., importance sampling weights normalized over all records), which is the key architectural limitation of \sn that our algorithms overcome. Since all workers execute the same procedure, we present both algorithms from the perspective of a single worker.

A direct consequence is the architectural property that motivates our design. Because each worker's sampling weights and threshold estimates come from batch-local statistics alone (Sections~\ref{sec:supgit} and~\ref{sec:gamcal}), neither algorithm needs a global pass over the proxy-score distribution before it can act. The first oracle delegations, and thus the first emitted results, become available after a single batch rather than after a full pass over the input, as \sn-based cascades require. On the 250K-row NYT join with a batch size of $4{,}096$, that reduces the number of rows processed before the first output by roughly $60\times$.

Within each batch, both algorithms sample a subset of records for oracle evaluation. The per-batch sampling budget is $k_t = \lfloor \rho \cdot |B_t| \rfloor$ oracle calls, where $\rho \in (0, 1]$ is the \emph{budget fraction}. Sampled records receive oracle labels that serve dual purposes: they inform the algorithm's estimates, and they provide correct classifications for the sampled records themselves.

Per-worker quality guarantees compose to a global guarantee on the same threshold $t_P$, with the confidence level degraded by a union bound. Both global precision and global recall are weighted averages of their per-worker counterparts, with weights given by per-worker positive predictions and per-worker positives respectively. As a result, $\text{Precision}_w \geq t_P$ holding deterministically for every worker implies global $\text{Precision} \geq t_P$, and likewise for recall. Probabilistically, if each of $W$ independent workers satisfies $\Pr[\text{Precision}_w \geq t_P] \geq 1 - \delta$, the union bound gives $\Pr[\text{global Precision} \geq t_P] \geq 1 - W\delta$. Setting $\delta = \delta_{\text{global}} / W$ at each worker therefore recovers any desired global confidence level. Appendix~\ref{app:dop} confirms empirically that quality remains stable as $W$ increases.

\section{\supgit: Iterative SUPG with Joint Targets}\label{sec:supgit}

\supgit operates in the streaming execution model formalized in Section~\ref{sec:problem}, replacing single-pass batch execution with \emph{iterative} threshold refinement where importance-sampled oracle labels accumulate across batches and drive progressively tighter estimates (Section~\ref{sec:supgit-sampling}). Within \sn's statistical framework, \supgit further introduces \emph{joint} precision-recall targeting using the two-threshold decision framework of Section~\ref{sec:problem}, refining $\tau_{\text{high}}$ and $\tau_{\text{low}}$ simultaneously rather than optimizing a single metric in isolation (Section~\ref{sec:supgit-iterative}). The combination of iterative execution and joint targeting raises challenges absent from single-pass, single-metric designs: Sections~\ref{sec:supgit-corrections}--\ref{sec:supgit-scope} present mechanisms for bounding sampling uncertainty in the corrected recall target, resolving conflicting precision-recall constraints, and preventing feedback loops in iterative threshold estimation.

\begin{figure}[t]
  \centering
  \includegraphics[width=\columnwidth]{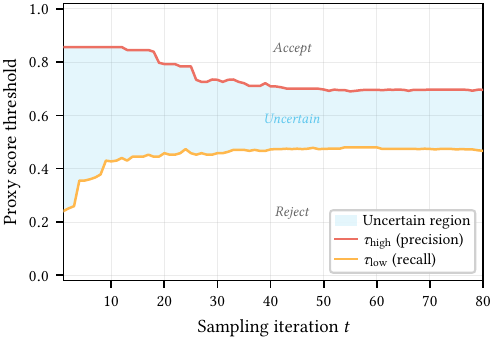}
  \caption{Threshold convergence on synthetic data with overlapping bimodal class distributions ($m = 5{,}000$ records, $k_t = 20$ samples per iteration, $t_R = t_P = 0.8$). The recall threshold $\tau_{\text{low}}$ (orange) rises as oracle samples accumulate. The precision threshold $\tau_{\text{high}}$ (red) descends as the confidence bound on precision tightens. The shaded uncertain region narrows accordingly, reducing oracle delegation.}
  \label{fig:threshold-convergence}
\end{figure}

Algorithm~\ref{alg:supgit} presents the complete procedure. The algorithm maintains two thresholds ($\tau_{\text{high}}$ for precision and $\tau_{\text{low}}$ for recall) that are iteratively refined as evidence grows over successive batches. In the first batch, thresholds are estimated from only $k_1$ oracle samples. As samples accumulate across subsequent batches, estimates converge and the uncertain region narrows (Figure~\ref{fig:threshold-convergence}).

\begin{algorithm}[t]
\caption{\supgit: Iterative SUPG with Joint Targets}
\label{alg:supgit}
\begin{algorithmic}[1]
\Require Dataset $\SD$, proxy model $A$, oracle $O$, targets $t_P, t_R$, failure probability $\delta$, budget fraction $\rho$, mixing coefficient $\eta$

\State $\SSe \gets \emptyset$ \Comment{Accumulated samples}
\State $\tau_{\text{low}} \gets 0, \tau_{\text{high}} \gets 1$ \Comment{Initial thresholds (wide uncertain region)}

\For{each batch $B_t \subseteq \SD$}
    \State Compute proxy scores $\{A(x_i)\}$ for $x_i \in B_t$
    \State $k_t \gets \lfloor \rho \cdot |B_t| \rfloor$ \Comment{Sampling budget}
    \State Compute weights $w_i$ via Eq.~\ref{eq:importance-weights}
    \State Sample $S_t$ ($k_t$ records w/o replacement, probabilities $\propto w$)
    \State Query oracle: $y_i \gets O(x_i)$ for $x_i \in S_t$
    \State $\SSe \gets \SSe \cup \{(A(x_i), y_i, \gamma_i) : x_i \in S_t\}$
    
    \State \Comment{Update thresholds}
    \State Compute $\hat{\tau}_{\text{low}}$ from weighted ROC curve
    \State Compute corrected target $t_R''$ via Eq.~\ref{eq:corrected-recall}, \ref{eq:target-clipping}
    \State $\tau_{\text{low}} \gets$ threshold achieving $t_R''$ recall
    \State $\tau_{\text{high}} \gets$ min threshold with LB precision $\geq t_P$
    
    \If{$\tau_{\text{high}} < \tau_{\text{low}}$} \Comment{Conflict resolution}
        \State $\tau_{\text{low}} \gets \tau_{\text{high}} \gets \tau_{\text{balanced}}$ via Eq.~\ref{eq:balanced-threshold}
    \EndIf
    
    \State \Comment{Classify non-sampled records in $B_t$}
    \State $\hat{y}_i \gets y_i$ for $x_i \in S_t$ \Comment{Sampled records use oracle labels}
    \For{each $x_i \in B_t \setminus S_t$}
        \If{$A(x_i) < \tau_{\text{low}}$}
            \State $\hat{y}_i \gets 0$ \Comment{Reject}
        \ElsIf{$A(x_i) \geq \tau_{\text{high}}$}
            \State $\hat{y}_i \gets 1$ \Comment{Accept}
        \Else
            \State $\hat{y}_i \gets O(x_i)$ \Comment{Delegate to oracle}
        \EndIf
    \EndFor
\EndFor
\end{algorithmic}
\end{algorithm}

\subsection{Importance Sampling with Defensive Mixing}\label{sec:supgit-sampling}

Accurate threshold estimation requires oracle labels spanning the proxy-score distribution, not only near the decision boundary. We use importance sampling to concentrate labeling on high-impact records while preserving support over the full batch through defensive mixing.

Given proxy scores $\{A(x_i)\}$ for the $m = |B_t|$ records in the current batch, we compute sampling weights:
\begin{equation}
    w_i = \eta \cdot \frac{\sqrt{A(x_i)}}{\sum_j \sqrt{A(x_j)}} + (1 - \eta) \cdot \frac{1}{m}
    \label{eq:importance-weights}
\end{equation}

where $\eta \in [0, 1]$ controls the importance--uniform tradeoff. The normalization $\sum_j \sqrt{A(x_j)}$ is computed over the current batch $B_t$, not the entire dataset, which is what permits streaming execution without the global coordination that \sn requires.

The first term follows \sn's importance sampling scheme where $w(x) \propto \sqrt{A(x)}$, the variance-optimal choice~\cite{neyman1934two, kang2020approximate} for estimating the weighted mean $\frac{1}{m}\sum_i A(x_i) y_i$. The second term is a uniform component that serves as \emph{defensive mixing}~\cite{owen2000safe}. It guarantees a minimum per-record sampling probability of $(1-\eta)/m$ and prevents a failure mode of pure importance sampling: if the proxy systematically assigns low scores to a subpopulation of positives, those records are missed entirely and downstream recall estimates become biased.

Unlike \sn, which samples with replacement, \supgit samples \emph{without replacement} from each batch, drawing $k_t$ records with probabilities proportional to $\{w_i\}$. Sampling without replacement reduces estimator variance by the finite population correction factor $(1 - k_t/m)$, a substantial improvement when batch sizes are modest relative to the sample budget.

To ensure that weighted statistics remain unbiased despite non-uniform sampling, each sampled record $x_i$ receives a Horvitz--Thompson~\cite{horvitz1952generalization} inverse-probability correction:
\begin{equation}
    \gamma_i = \frac{1/m}{w_i}
    \label{eq:correction-factor}
\end{equation}

\subsection{Iterative Threshold Refinement}\label{sec:supgit-iterative}

The threshold refinement procedure uses the confidence bounds inherited from \sn~\cite{kang2020approximate}. For a sample mean $\mu$ with standard deviation $\sigma$ computed from $s$ samples, the upper and lower bounds at confidence level $1 - \delta$ are:
\begin{align}
    \text{UB}(\mu, \sigma, s, \delta) &= \mu + \frac{\sigma}{\sqrt{s}} \sqrt{2 \ln(1/\delta)} \label{eq:ub} \\
    \text{LB}(\mu, \sigma, s, \delta) &= \mu - \frac{\sigma}{\sqrt{s}} \sqrt{2 \ln(1/\delta)} \label{eq:lb}
\end{align}
where $\sqrt{2 \ln(1/\delta)}$ is a conservative upper bound on the Gaussian quantile $\Phi^{-1}(1-\delta)$~\cite{wasserman2013all}. The bounds rely on the same central-limit approximation as \sn, so the precision and recall targets hold with high probability under that approximation rather than as finite-sample certainties. Section~\ref{sec:exp-reliability} confirms that they hold empirically at $\delta = 0.2$.

\supgit processes data in batches, refining thresholds after each batch. Let $\SSe^{(t)} = \{(A(x_i), y_i, \gamma_i)\}$ denote the accumulated sample after processing $t$ batches, where $y_i$ is the oracle label and $\gamma_i$ is the correction factor from Eq.~\ref{eq:correction-factor}.

\paragraph{Recall threshold ($\tau_{\text{low}}$).} We compute the threshold that achieves the target recall using the weighted ROC curve. Given samples sorted by descending proxy score, the weighted true positive rate at threshold $\tau$ is:
\begin{equation}
    \text{TPR}(\tau) = \frac{\sum_{i: A(x_i) \geq \tau} \gamma_i \cdot y_i}{\sum_i \gamma_i \cdot y_i}
\end{equation}

Since $\text{TPR}(\tau)$ is monotonically non-increasing in $\tau$, the recall threshold $\hat{\tau}_{\text{low}}$ is the \emph{largest} threshold satisfying $\text{TPR}(\tau) \geq t_R$, i.e., the most selective threshold that still achieves the user-specified recall target.

\paragraph{Precision threshold ($\tau_{\text{high}}$).} For precision, we compute cumulative statistics over samples sorted by descending proxy score. At each candidate threshold $\tau$, let:
\begin{align}
    \mu(\tau) &= \frac{\sum_{i: A(x_i) \geq \tau} y_i}{|\{i: A(x_i) \geq \tau\}|} \\
    \sigma(\tau) &= \sqrt{\mu(\tau)(1 - \mu(\tau))}
\end{align}

Note that $\mu(\tau)$ uses raw oracle labels $y_i$ without correction factors $\gamma_i$, unlike the recall computation. The omission follows the original \sn design~\cite{kang2020approximate} because precision is a ratio where both numerator and denominator sum over the same subpopulation (records above $\tau$), so the Horvitz--Thompson corrections approximately cancel.

The precision threshold $\tau_{\text{high}}$ is the minimum threshold where the statistical lower bound exceeds the precision target:
\begin{equation}
    \tau_{\text{high}} = \min\{\tau : \text{LB}(\mu(\tau), \sigma(\tau), s_\tau, \delta') \geq t_P\}
\end{equation}
where $s_\tau = |\{i: A(x_i) \geq \tau\}|$ is the sample count above threshold $\tau$ and $\delta' = \delta / |\SSe^{(t)}|$ applies a Bonferroni correction, since each unique proxy score in the accumulated sample defines a candidate threshold.

\subsection{Statistical Correction with Target Clipping}\label{sec:supgit-corrections}

The initial recall threshold $\hat{\tau}_{\text{low}}$ is computed from a finite sample and may not achieve the target recall on the full dataset. Following \sn~\cite{kang2020approximate}, we apply a \emph{corrected recall target}, inflated to account for sampling uncertainty.

For each record $x_i \in \SSe^{(t)}$, define indicator-weighted statistics over the full accumulated sample:
\begin{align}
    Z_{1,i} &= \gamma_i \cdot y_i \cdot \mathbf{1}[A(x_i) \geq \hat{\tau}_{\text{low}}] \\
    Z_{2,i} &= \gamma_i \cdot y_i \cdot \mathbf{1}[A(x_i) < \hat{\tau}_{\text{low}}]
\end{align}

Both sequences have $s = |\SSe^{(t)}|$ elements, with records on the opposite side of the threshold contributing zero. Sample recall decomposes as $\bar{Z}_1 / (\bar{Z}_1 + \bar{Z}_2)$, which is increasing in $\bar{Z}_1$ and decreasing in $\bar{Z}_2$. An upper bound on the sample recall at the true optimal threshold is therefore obtained by replacing these with their respective confidence bounds:
\begin{equation}
    t_R' = \frac{\text{UB}(\bar{Z}_1, \sigma_{Z_1}, s, \delta/2)}{\text{UB}(\bar{Z}_1, \sigma_{Z_1}, s, \delta/2) + \text{LB}(\bar{Z}_2, \sigma_{Z_2}, s, \delta/2)}
    \label{eq:corrected-recall}
\end{equation}

Since $t_R' \geq t_R$ in general, the corrected target requires a \emph{lower} (more inclusive) recall threshold, which is more conservative: fewer records are rejected outright, reducing the risk of missed positives.

However, the correction in Equation~\ref{eq:corrected-recall} can produce extreme values when sample sizes are small or distributions are skewed. \supgit extends \sn by applying \emph{target clipping} to bound the corrected target:
\begin{equation}
    t_R'' = \text{clip}(t_R', t_R, t_R + \Delta)
    \label{eq:target-clipping}
\end{equation}
where $\Delta$ is a small constant. Clipping prevents over-correction that would unnecessarily increase the delegation rate, while still ensuring the corrected target remains at least as conservative as the original.

The final recall threshold is computed using the clipped target:
\begin{equation}
    \tau_{\text{low}} = \max\{\tau : \text{TPR}(\tau) \geq t_R''\}
\end{equation}

\subsection{Threshold Conflict Resolution}\label{sec:supgit-conflicts}

Joint precision-recall targeting can produce conflicting constraints where $\tau_{\text{high}} < \tau_{\text{low}}$, creating an invalid configuration in which the accept region falls below the reject region. Such conflicts arise when the proxy is poorly calibrated or the targets are jointly difficult to achieve with the available sample.

When a conflict is detected, \supgit collapses to a single \emph{balanced threshold} that best matches the user's desired precision-recall ratio:
\begin{equation}
    \tau_{\text{low}} = \tau_{\text{high}} = \tau_{\text{balanced}} = \argmin_\tau \left| \frac{\text{Recall}(\tau)}{\text{Precision}(\tau)} - \frac{t_R}{t_P} \right|
    \label{eq:balanced-threshold}
\end{equation}
where Recall and Precision are computed from the precision-recall curve over the accumulated sample. Note that collapsing to a single threshold eliminates the uncertain region entirely, so no further records are delegated to the oracle in this batch (the cascade reduces to a simple threshold classifier). The choice gracefully relaxes both targets in the user-specified ratio $t_R/t_P$ rather than arbitrarily sacrificing one, and the collapse is not permanent: subsequent batches re-estimate $\tau_{\text{low}}$ and $\tau_{\text{high}}$ afresh from the accumulated sample, restoring the two-threshold regime once additional oracle labels make the targets jointly feasible.

\subsection{Expanded Sampling Scope}\label{sec:supgit-scope}

The two-threshold decision framework introduced in Section~\ref{sec:problem} partitions records into accept, reject, and uncertain regions, a structure absent from \sn, which uses a single threshold and samples from the entire dataset in one pass. In the iterative, streaming setting where thresholds evolve over successive batches, a natural approach would be to restrict sampling to the uncertain region, focusing the oracle budget on records that have not yet been confidently classified. However, restricting sampling creates a feedback loop: inaccurate initial thresholds narrow the uncertain region, confining sampling to a subset of the score distribution, which in turn prevents the algorithm from correcting its thresholds.

\supgit avoids this by sampling from \emph{all remaining records} in each batch, regardless of whether they fall in the accept, reject, or uncertain region:
\begin{equation}
    \text{Sampling pool} = \{x_i : x_i \in B_t,\; x_i \notin \SSe^{(t-1)}\}
\end{equation}
The expanded sampling scope enables estimation of the score distribution across all three regions, helps detect proxy miscalibration through high-confidence errors, and provides more robust threshold refinement when initial estimates are poor.

\subsection{Handling Uncertain Records}\label{sec:supgit-residual}

After the sampling budget is exhausted, records in the uncertain region ($\tau_{\text{low}} \leq A(x) < \tau_{\text{high}}$) require a decision. \supgit supports two strategies, chosen based on whether the application prioritizes quality or cost.

\paragraph{Oracle delegation.} The default strategy (as shown in Algorithm~\ref{alg:supgit}) sends all remaining uncertain records to the oracle, guaranteeing correctness at the expense of additional oracle calls. In practice, the uncertain region typically shrinks as samples accumulate across batches, limiting the delegation rate.

\paragraph{Threshold-based fallback.} When the oracle budget is constrained, an alternative strategy applies a single threshold to uncertain records, classifying them by proxy score alone:
\begin{equation}
    \tau_{\text{mid}} = \argmax_\tau F_1(\tau)
\end{equation}
computed over the accumulated sample. The fallback eliminates oracle delegation for uncertain records at the cost of potential quality degradation.

\section{\gamcal: Calibration-Based Cascade}\label{sec:gamcal}

\gamcal takes a different approach from \supgit by replacing \emph{statistical threshold estimation} with \emph{learned calibration}. The key insight is that if we can learn a monotone function $g: [0,1] \to [0,1]$ mapping raw proxy scores to calibrated probabilities $g(A(x)) \approx P(y=1 \mid A(x))$, then for any threshold pair $(\tau_{\text{low}}, \tau_{\text{high}})$ we can \emph{predict} the expected precision, recall, and delegation rate in closed form, without additional oracle calls. The resulting decoupling of threshold optimization from oracle evaluation enables direct numerical optimization of a cost-quality objective in the streaming setting, where past batches cannot be revisited.

Like \supgit, \gamcal operates in the streaming execution model of Section~\ref{sec:problem}: each worker processes batches independently, accumulates oracle samples, and refines its local thresholds. However, where \supgit uses oracle labels to tighten confidence bounds on empirical metrics, \gamcal uses them to train a Generalized Additive Model (GAM) that provides calibrated probability estimates with uncertainty quantification. Thresholds are then optimized against a continuous cost-quality tradeoff rather than fixed precision/recall targets, eliminating the need for users to specify targets \emph{a priori}.

Algorithm~\ref{alg:gamcal} presents the complete procedure. The remainder of this section motivates the approach (Section~\ref{sec:gamcal-motivation}), describes GAM-based calibration (Section~\ref{sec:gamcal-calibration}), explains uncertainty-aware routing via random quantiles (Section~\ref{sec:gamcal-stochastic}), presents the threshold optimization objective (Section~\ref{sec:gamcal-optimization}), and discusses the adaptive retraining schedule (Section~\ref{sec:gamcal-retraining}) and oracle sampling strategy (Section~\ref{sec:gamcal-sampling}).

\begin{algorithm}[t]
\caption{\gamcal: GAM-Calibrated Cascade}
\label{alg:gamcal}
\begin{algorithmic}[1]
\Require Dataset $\SD$, proxy $A$, oracle $O$, tradeoff $\alpha$, $F_\beta$ weight $\beta$, budget fraction $\rho$, smoothing $\lambda$, min samples $n_{\min}$

\State $\SSe \gets \emptyset$, $n_{\text{last}} \gets 0$, $g \gets \text{identity}$ \Comment{Initialize}
\State $\tau_{\text{low}} \gets 0$, $\tau_{\text{high}} \gets 1$ \Comment{Initial thresholds (wide uncertain region)}
\State Compute proxy scores $\{A(x_i)\}$ and sample $q_i \sim \text{Uniform}(0, 1)$ for all $x_i \in \SD$

\For{each batch $B_t \subseteq \SD$}
    \State Compute $\tilde{g}(A(x_i), q_i)$ via Eq.~\ref{eq:calibrated-score} for each $x_i \in B_t$
    \State $U_t \gets \{i : \tau_{\text{low}} \leq \tilde{g}(A(x_i), q_i) < \tau_{\text{high}}\}$ \Comment{Uncertain}
    \State $k_t \gets \min(\lfloor \rho \cdot |B_t| \rfloor, |U_t|)$
    
    \State Sample $S_t \sim \text{Uniform}(U_t, k_t)$ without replacement
    \State Query oracle: $y_i \gets O(x_i)$ for $x_i \in S_t$
    \State $\SSe \gets \SSe \cup \{(A(x_i), y_i) : x_i \in S_t\}$
    
    \If{$|\SSe| \geq 2 \cdot n_{\text{last}}$ \textbf{and} $\min\bigl(|\{i \in \SSe : y_i\!=\!1\}|,\; |\{i \in \SSe : y_i\!=\!0\}|\bigr) \geq n_{\min}$}
        \State Train GAM $g$ on $\SSe$ via Eq.~\ref{eq:gam-likelihood}
        \State Recompute $\tilde{g}(A(x_i), q_i)$ for all $x_i \in \SD$
        \State Optimize $(\tau_{\text{low}}, \tau_{\text{high}})$ via Eq.~\ref{eq:gamcal-obj-concrete}
        \State $n_{\text{last}} \gets |\SSe|$
    \EndIf
    
    \State \Comment{Classify non-sampled records in $B_t$}
    \State $\hat{y}_i \gets y_i$ for $x_i \in S_t$ \Comment{Sampled records use oracle labels}
    \For{each $x_i \in B_t \setminus S_t$}
        \If{$\tilde{g}(A(x_i), q_i) < \tau_{\text{low}}$}
            \State $\hat{y}_i \gets 0$ \Comment{Reject}
        \ElsIf{$\tilde{g}(A(x_i), q_i) \geq \tau_{\text{high}}$}
            \State $\hat{y}_i \gets 1$ \Comment{Accept}
        \Else
            \State $\hat{y}_i \gets \tilde{g}(A(x_i), q_i) \geq 0.5$ \Comment{Fallback}
        \EndIf
    \EndFor
\EndFor
\end{algorithmic}
\end{algorithm}

\subsection{From Statistical Estimation to Learned Calibration}\label{sec:gamcal-motivation}

While \supgit addresses \sn's limitations around streaming execution and joint precision-recall targeting, it inherits deeper assumptions from the \sn framework that limit its effectiveness. Three observations motivate the shift to a calibration-based approach.

\textbf{Proxy miscalibration.} Both \sn and \supgit assume that proxy scores are approximately calibrated, i.e., that $A(x) \approx P(y=1 \mid x)$. The importance sampling weights $w(x) \propto \sqrt{A(x)}$ are variance-optimal only under calibration~\cite{kang2020approximate}, and the confidence bounds on precision and recall depend on the proxy scores' fidelity as probability estimates. In practice, proxy models, particularly small LLMs and embedding-based classifiers, are often poorly calibrated~\cite{guo2017calibration}. A score of 0.7 may correspond to a true positive rate anywhere from 0.4 to 0.95 depending on the dataset and predicate. Poor calibration renders \supgit's confidence bounds unreliable and produces conservative thresholds with unnecessarily high delegation rates.

\textbf{Limited generalization from oracle labels.} In \supgit, each oracle label contributes to threshold estimation only through running statistics (weighted means and variances) computed over the score region it falls in. No mechanism allows labels at one proxy score level to inform predictions at other levels. A smooth calibration model provides exactly such generalization: labels at scattered score levels jointly constrain a function that interpolates across the entire score range. In the streaming setting, where early batches yield few oracle labels, such interpolation extracts substantially more information per label than raw statistical estimation.

\textbf{Continuous cost-quality tradeoff.} Beyond these technical concerns, the target-based formulation creates practical challenges. Specifying precision and recall targets requires \emph{a priori} knowledge of dataset difficulty. Identical targets produce vastly different delegation rates across workloads, and target-based methods exhibit binary success/failure behavior with no graceful degradation. A single parameter $\alpha \in [0,1]$ governs the tradeoff between classification error and oracle cost (Eq.~\ref{eq:gamcal-objective}). Replacing fixed targets with this continuous objective allows the cascade to adapt automatically to the proxy's intrinsic accuracy.

Taken together, these observations motivate \gamcal's three-stage pipeline: \emph{calibrate} proxy scores using a GAM trained on oracle samples, \emph{predict} expected quality metrics for any threshold configuration from the calibrated model, and \emph{optimize} thresholds by minimizing the cost-quality objective via numerical optimization.

\subsection{GAM-Based Probability Calibration}\label{sec:gamcal-calibration}

At the calibrate stage, \gamcal learns a monotone function $g: [0,1] \to [0,1]$ mapping raw proxy scores $s = A(x)$ to calibrated probabilities $g(s) \approx P(y=1 \mid A(x) = s)$. Several calibration methods have been proposed: Platt scaling fits a logistic function to raw scores~\cite{platt1999probabilistic}, isotonic regression fits a non-parametric monotone map~\cite{zadrozny2002transforming}, and temperature scaling adjusts a single parameter~\cite{guo2017calibration}. However, none of these methods fully satisfies the requirements of the cascade setting.

Platt scaling assumes a linear relationship in log-odds space, which is too restrictive when the proxy model's miscalibration is nonlinear. Isotonic regression makes no smoothness assumptions and tends to overfit when oracle labels are scarce (precisely the regime encountered in early streaming batches). Neither method provides calibrated uncertainty estimates, which are needed for the uncertainty-aware routing described in Section~\ref{sec:gamcal-stochastic}.

Generalized Additive Models (GAMs)~\cite{hastie1986generalized} occupy a favorable middle ground. A logistic GAM models the calibration function through a smooth spline in log-odds space:
\begin{equation}
    \log \frac{g(s)}{1 - g(s)} = f(s)
    \label{eq:gam-logit}
\end{equation}
where $s = A(x)$ is the raw proxy score and $f$ is a smooth function represented by a cubic B-spline basis. The model is fit on accumulated oracle samples $\SSe = \{(A(x_i), y_i)\}$ by maximizing the penalized log-likelihood:
\begin{equation}
\begin{aligned}
    \mathcal{L}(f) ={} & \sum_{i=1}^{|\SSe|} \left[ y_i \log g(A(x_i)) + (1-y_i) \log(1 - g(A(x_i))) \right] \\
    & - \lambda \int_{0}^{1} \bigl(f''(s)\bigr)^2 \, ds
\end{aligned}
    \label{eq:gam-likelihood}
\end{equation}

The roughness penalty $\lambda \int (f''(s))^2\, ds$ controls the bias-variance tradeoff: large $\lambda$ produces smoother calibration curves that generalize from few samples (important in early batches), while small $\lambda$ allows the model to capture fine-grained calibration structure as oracle labels accumulate. Monotonicity ($g'(s) \geq 0$) is enforced through linear inequality constraints on the spline coefficients during optimization and guarantees that higher proxy scores always map to higher calibrated probabilities.

\begin{figure*}[t]
  \centering
  \includegraphics[width=0.83\textwidth]{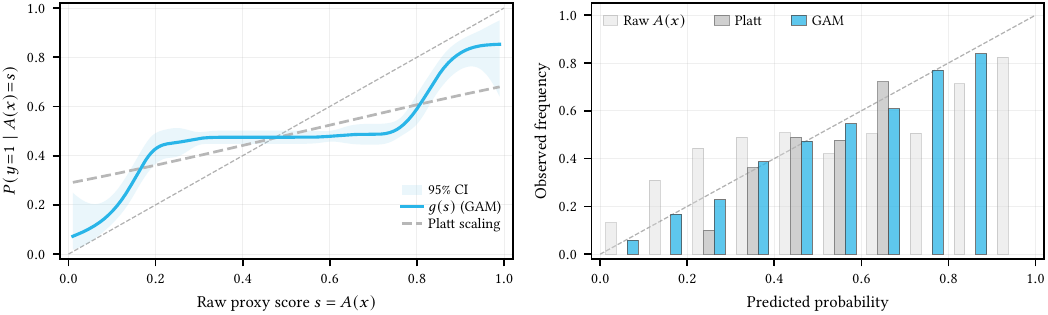}
  \caption{GAM-based calibration on synthetic data with nonlinear miscalibration ($n = 3{,}000$, $\lambda = 0.6$). \emph{Left:} The GAM calibration curve $g(s)$ (blue) captures the S-shaped departure from the diagonal that Platt scaling (gray dashed) cannot represent. The shaded band shows the 95\% confidence interval from the GAM posterior. \emph{Right:} Reliability diagram comparing raw, Platt-calibrated, and GAM-calibrated scores. GAM calibration reduces expected calibration error from $0.140$ (raw) to $0.005$, compared to $0.047$ for Platt scaling.}
  \label{fig:calibration}
\end{figure*}

GAMs combine the strengths that the alternatives lack: the spline basis captures nonlinear miscalibration patterns that Platt scaling misses (Figure~\ref{fig:calibration}), while the smoothness penalty prevents overfitting when oracle samples are limited, unlike isotonic regression. For the cascade setting, the penalized likelihood framework yields a posterior approximation over $f$: both a mean prediction $\hat{f}(s)$ and a standard error $\text{se}(s)$ at each score level. Given a quantile parameter $q \in [0,1]$, the stochastic calibrated score is:
\begin{equation}
    \tilde{g}(s, q) = \text{logit}^{-1}\!\left(\hat{f}(s) + \Phi^{-1}(q) \cdot \text{se}(s)\right)
    \label{eq:calibrated-score}
\end{equation}
where $\Phi^{-1}$ is the standard normal quantile function. Operating in log-odds space and applying the sigmoid ensures that $\tilde{g}(s,q) \in (0,1)$ for all $q$.

Our primary implementation uses constrained cubic splines via the \texttt{pyGAM} library. Appendix~\ref{app:calibration-variants} describes two alternative implementations (a regularized logistic regression variant with a Platt scaling prior and a bootstrap ensemble) that offer different tradeoffs between monotonicity guarantees, computational cost, and uncertainty robustness.

\subsection{Uncertainty-Aware Routing via Random Quantiles}\label{sec:gamcal-stochastic}

A key design choice in \gamcal is that routing decisions are \emph{stochastic} rather than deterministic. The stochastic calibrated score $\tilde{g}(s, q)$ (Eq.~\ref{eq:calibrated-score}) incorporates a random quantile $q_i \sim \text{Uniform}(0, 1)$ sampled independently for each record $x_i$. The quantile $q_i$ is drawn once and held fixed for the lifetime of the record, ensuring consistent routing across retraining events. For reproducibility, each $q_i$ is derived deterministically from a stable record identifier and a query-level seed rather than sampled from a global stream. Re-executing the same query then reproduces the same routing and output, and because the derivation is per-record it stays consistent across parallel workers without coordination.

Conceptually, the mechanism is a form of \emph{posterior sampling}~\cite{russo2018tutorial} applied to the calibration model: rather than routing each record based on the posterior mean $\hat{f}(s)$, \gamcal draws a sample from the approximate posterior over the calibrated log-odds and routes based on that draw. The standard error $\text{se}(s)$ from the GAM controls the spread: when calibration uncertainty is high, the sampled scores are dispersed. When calibration is confident, the samples cluster near the mean. Consequently, records whose proxy scores fall in poorly calibrated regions receive more dispersed calibrated scores and are more likely to land in the uncertain region $[\tau_{\text{low}}, \tau_{\text{high}})$, directing oracle budget toward the records where calibration is least reliable.

\paragraph{Soft decision boundaries.} Deterministic thresholds create sharp boundaries: records with calibrated scores just above $\tau_{\text{high}}$ are always accepted, while those just below are always delegated. When many records cluster near a threshold, small estimation errors produce large swings in the delegation rate. Stochastic scoring replaces each sharp boundary with a smooth transition zone, where the probability of delegation decreases continuously as the calibrated score moves away from the threshold. The width of the transition zone adapts automatically: it is wide where $\text{se}(s)$ is large (few oracle labels nearby) and narrow where $\text{se}(s)$ is small (many oracle labels provide precise calibration).

\paragraph{Exploration for calibration improvement.} Stochastic routing ensures that records near decision boundaries are occasionally delegated to the oracle even when the current model would confidently classify them. The resulting oracle labels provide training data in score regions that deterministic routing would never query. The calibration model can then refine its estimates in subsequent retraining events (Section~\ref{sec:gamcal-retraining}). The mechanism mirrors Thompson sampling~\cite{thompson1933likelihood, russo2018tutorial}, where actions (routing decisions) are randomized according to the posterior probability of being optimal. Exploration is therefore proportional to uncertainty and vanishes as the model converges. Unlike \supgit, which achieves exploration through a separate design choice (expanded sampling scope, Section~\ref{sec:supgit-scope}), \gamcal obtains exploration as a natural byproduct of posterior sampling, requiring no additional mechanism or tuning.

Uncertainty-directed delegation, soft boundaries, and exploration all arise from a single mechanism (per-record random quantile draws) with no additional parameters beyond the quantile distribution itself. The implicit exploration is precisely what allows \gamcal to use the simpler oracle sampling strategy described in Section~\ref{sec:gamcal-sampling}: because stochastic routing already diversifies the records entering the uncertain region, there is no need for importance sampling or expanded sampling scope.

\subsection{Direct Threshold Optimization}\label{sec:gamcal-optimization}

Once the GAM provides calibrated probabilities, the cascade can \emph{predict} the expected quality of any threshold configuration without additional oracle queries. For each record $x_i$, let $\tilde{g}_i = \tilde{g}(A(x_i), q_i)$ denote the stochastic calibrated score from Eq.~\ref{eq:calibrated-score}. Under the two-threshold framework of Section~\ref{sec:problem}, thresholds $(\tau_{\text{low}}, \tau_{\text{high}})$ partition records into reject, accept, and uncertain regions. Because $\tilde{g}_i$ estimates the probability that record $x_i$ is a true positive, each region's expected contribution to the confusion matrix can be computed in closed form:

\begin{itemize}
    \item \textbf{Reject region} ($\tilde{g}_i < \tau_{\text{low}}$): Each rejected record has probability $\tilde{g}_i$ of being a true positive, yielding $\sum_{i:\, \tilde{g}_i < \tau_{\text{low}}} \tilde{g}_i$ expected false negatives.
    
    \item \textbf{Accept region} ($\tilde{g}_i \geq \tau_{\text{high}}$): Accepted records contribute $\sum \tilde{g}_i$ expected true positives and $\sum (1 - \tilde{g}_i)$ expected false positives.
    
    \item \textbf{Uncertain region} ($\tau_{\text{low}} \leq \tilde{g}_i < \tau_{\text{high}}$): Oracle-delegated records are classified correctly, contributing $\sum \tilde{g}_i$ expected true positives and zero classification error.
\end{itemize}

The uncertain region contributes no classification error, only oracle cost. Aggregating across regions:
\begin{align}
    \ex[\text{TP}] &= \sum_{i:\, \tilde{g}_i \geq \tau_{\text{low}}} \tilde{g}_i \label{eq:expected-tp} \\
    \ex[\text{FP}] &= \sum_{i:\, \tilde{g}_i \geq \tau_{\text{high}}} (1 - \tilde{g}_i) \label{eq:expected-fp} \\
    \ex[\text{FN}] &= \sum_{i:\, \tilde{g}_i < \tau_{\text{low}}} \tilde{g}_i \label{eq:expected-fn}
\end{align}
where $\ex[\text{TP}]$ combines both the accept and uncertain regions (both contribute true positives, the former by proxy prediction, the latter by oracle evaluation). The expected $F_\beta$ score follows directly:
\begin{equation}
    \ex[F_\beta] = \frac{(1 + \beta^2) \cdot \ex[\text{TP}]}{(1 + \beta^2) \cdot \ex[\text{TP}] + \ex[\text{FN}] + \beta^2 \cdot \ex[\text{FP}]}
    \label{eq:expected-fbeta}
\end{equation}
with $\ex[F_\beta] = 0$ when $\ex[\text{TP}] = 0$ (i.e., all records fall in the reject region).

The objective from Eq.~\ref{eq:gamcal-objective} is instantiated as:
\begin{equation}
    \min_{\tau_{\text{low}}, \tau_{\text{high}}} \; \alpha \cdot \underbrace{\frac{1 - \ex[F_\beta(\tau_{\text{low}}, \tau_{\text{high}})]}{1 - \ex[F_\beta(0.5, 0.5)]}}_{\text{normalized error}} + (1 - \alpha) \cdot \underbrace{\frac{|\{i : \tau_{\text{low}} \leq \tilde{g}_i < \tau_{\text{high}}\}|}{n}}_{\text{delegation rate}}
    \label{eq:gamcal-obj-concrete}
\end{equation}
The error term is normalized by the error of a \emph{no-delegation baseline} that classifies all records by proxy at threshold $0.5$ (i.e., $\tau_{\text{low}} = \tau_{\text{high}} = 0.5$). The delegation rate lies in $[0, 1]$ and the normalized error equals $1$ at the no-delegation baseline, so $\alpha$ interpolates meaningfully between the two objectives. Concretely, higher $\alpha$ places more weight on quality and widens the uncertain region, while lower $\alpha$ narrows the region to minimize oracle calls.

The $F_\beta$ computation assumes full delegation of the uncertain region. In practice, the budget fraction $\rho$ limits oracle calls to $\lfloor \rho \cdot |B_t| \rfloor$ per batch, and any excess uncertain records are classified using the calibrated score at threshold $0.5$ (Algorithm~\ref{alg:gamcal}). The objective thus models \emph{idealized} quality, but as the calibration model improves and thresholds converge, the uncertain region typically shrinks below the budget, closing the gap between predicted and realized $F_\beta$.

Optimization is complicated by the objective's piecewise-constant structure: because the expected confusion matrix terms are discrete sums over threshold-defined sets, the objective jumps discontinuously whenever a threshold crosses an individual calibrated score and is flat between successive scores. Gradient-based methods are therefore inapplicable. We optimize using differential evolution~\cite{storn1997differential}, a gradient-free global optimizer well-suited to piecewise-constant objectives. To enforce the constraint $\tau_{\text{low}} \leq \tau_{\text{high}}$, we reparameterize the search space as $(y_1, y_2) \in [0,1]^2$ with:
\begin{equation}
    \tau_{\text{low}} = y_1, \quad \tau_{\text{high}} = y_1 + (1 - y_1) \cdot y_2
    \label{eq:reparameterization}
\end{equation}
Here $y_1$ directly sets the lower threshold, while $y_2$ controls the gap as a fraction of the remaining range $[y_1, 1]$: $y_2 = 0$ collapses the uncertain region (no delegation), while $y_2 = 1$ places $\tau_{\text{high}} = 1$ (all non-rejected records are delegated).

\subsection{Adaptive Retraining Schedule}\label{sec:gamcal-retraining}

As oracle samples accumulate, \gamcal must periodically update the calibration model and thresholds. Each update incurs a computational cost, since it requires GAM fitting followed by threshold re-optimization via differential evolution (Section~\ref{sec:gamcal-optimization}), so retraining at every batch is wasteful.

\gamcal adopts a \emph{doubling schedule}~\cite{cesa2006prediction}, a standard technique in online learning: retrain when the accumulated sample size has at least doubled since the last training event, i.e., when $|\SSe| \geq 2 n_{\text{last}}$, where $n_{\text{last}}$ is the sample count at the previous training. For $n$ total oracle samples, the doubling schedule bounds the number of retraining events to $O(\log n)$, while ensuring that each successive model is trained on at least twice as much data as its predecessor.

Each retraining event triggers two operations: fitting the GAM on the full accumulated sample $\SSe$ (updating both the calibrated probabilities $\hat{f}(s)$ and the standard errors $\text{se}(s)$), and re-optimizing the thresholds $(\tau_{\text{low}}, \tau_{\text{high}})$ via Eq.~\ref{eq:gamcal-obj-concrete} using the updated calibrated scores. Both operations run on CPU over the accumulated oracle sample rather than the full table and, triggered only $O(\log n)$ times, add negligible end-to-end cost: each retraining is orders of magnitude cheaper than the oracle LLM calls issued in a single batch, so wall-clock cost tracks the delegation rate rather than the retraining schedule.

\begin{figure}[t]
  \centering
  \includegraphics[width=\columnwidth]{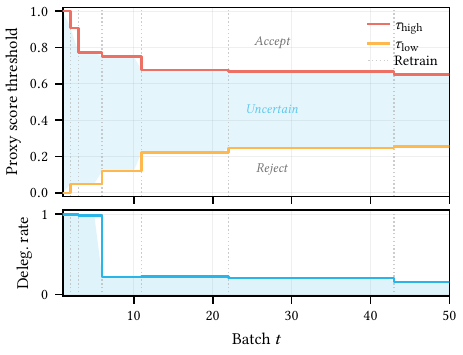}
  \caption{\gamcal threshold convergence on synthetic bimodal data ($m = 10{,}000$, batch size $200$, $\rho = 0.03$, $\alpha = 0.35$). \emph{Top:} Thresholds $\tau_{\text{low}}$ (orange) and $\tau_{\text{high}}$ (red) narrow in discrete steps at retraining events (dotted lines) on the doubling schedule. The shaded region marks the uncertain interval. \emph{Bottom:} Delegation rate drops from $1.0$ during the cold-start phase to approximately $0.15$ as calibration improves. Each retraining event further reduces delegation.}
  \label{fig:gamcal-convergence}
\end{figure}

To prevent overfitting during early execution, \gamcal defers the first training until both classes have accumulated a minimum number of samples:
\begin{equation}
    \min\left( |\{i \in \SSe : y_i = 0\}|, \; |\{i \in \SSe : y_i = 1\}| \right) \geq n_{\min}
\end{equation}
Before this condition is met, \gamcal operates in a \emph{cold-start phase} with default thresholds $\tau_{\text{low}} = 0$, $\tau_{\text{high}} = 1$ and identity calibration $g = \text{id}$. Under these defaults all records fall in the uncertain region, so \gamcal delegates every sampled record to the oracle. The resulting full-delegation strategy is conservative but maximizes information gain for the calibration model. For balanced datasets the cold-start phase typically ends within the first few batches. For highly imbalanced predicates, reaching $n_{\min}$ samples of the minority class may require more batches, which can be mitigated by increasing the budget fraction $\rho$. Figure~\ref{fig:gamcal-convergence} illustrates these dynamics on synthetic data.

\subsection{Oracle Sampling Strategy}\label{sec:gamcal-sampling}

\supgit uses importance sampling from all remaining records in each batch (Section~\ref{sec:supgit-scope}) to avoid feedback loops where inaccurate thresholds restrict the sampling pool. \gamcal takes a different approach: oracle labels are drawn by \emph{uniform random sampling without replacement} from the uncertain region:
\begin{align*}
    S_t &\sim \text{Uniform}(\{i \in B_t : \tau_{\text{low}} \leq \tilde{g}_i < \tau_{\text{high}}\}), \\
    |S_t| &= \min(\lfloor \rho \cdot |B_t| \rfloor,\; |U_t|)
\end{align*}
where $\rho$ is the budget fraction and $|U_t|$ is the number of uncertain records in batch $B_t$.

Expanded sampling is unnecessary because \gamcal's stochastic routing mechanism (Section~\ref{sec:gamcal-stochastic}) already ensures that records with high calibration uncertainty are drawn into the uncertain region, even if deterministic thresholds would place them in the accept or reject regions. The random quantiles provide implicit exploration across the score distribution. Moreover, the GAM's smooth spline basis generalizes from oracle labels in the uncertain region to predictions across the full score range (as argued in Section~\ref{sec:gamcal-motivation}), so sampling from a restricted score interval still informs calibration globally.

Uniform sampling also offers a practical advantage over importance sampling: all oracle labels contribute equally to the GAM training data and require no Horvitz--Thompson correction factors~\cite{horvitz1952generalization}. As thresholds converge and the uncertain region narrows, the sampling budget is naturally redirected: fewer records require oracle evaluation, reducing the effective delegation rate without explicit budget management.

\section{Experimental Evaluation}\label{sec:experiments}

\begin{table*}[t]
\begin{minipage}[t]{0.48\textwidth}
\centering
\small
\caption{Dataset characteristics. Six benchmarks spanning classification, filtering, and join operators with diverse proxy quality.}
\label{tab:datasets}
\begin{tabular}{llrrrr}
\toprule
Dataset & Task & Rows & Pos\% & Proxy F1 & ECE \\
\midrule
MMLU    & \texttt{AI\_CLASSIFY} & 5{,}000   & 71\%   & 0.817 & 0.110 \\
BoolQ   & \texttt{AI\_FILTER}   & 12{,}697  & 79\%   & 0.823 & 0.211 \\
IMDB    & \texttt{AI\_FILTER}   & 50{,}000  & 19\%   & 0.382 & 0.452 \\
ArXiv   & \texttt{AI\_FILTER}   & 56{,}181  & 8.5\%  & 0.528 & 0.050 \\
SST-2   & \texttt{AI\_FILTER}   & 68{,}221  & 46\%   & 0.819 & 0.095 \\
NYT     & \texttt{AI\_JOIN}     & 250{,}000 & 0.9\%  & 0.209 & 0.067 \\
\bottomrule
\end{tabular}
\end{minipage}%
\hfill
\begin{minipage}[t]{0.48\textwidth}
\centering
\small
\caption{Best $F_1$ operating point per algorithm. Each cascade cell shows $F_1$ score and delegation rate $d$. Bold marks the best cascade algorithm per dataset.}
\label{tab:best-f1}
\resizebox{\linewidth}{!}{%
\begin{tabular}{lccccc}
\toprule
Dataset & Proxy-only & \sn & \supgsp & \supgit & \gamcal \\
\midrule
ArXiv & $0.528$ & $0.541$ & $0.979$~(74\%) & $\mathbf{0.992}$~(86\%) & $0.973$~(69\%) \\
BoolQ & $0.823$ & $0.899$ & $0.980$~(82\%) & $0.990$~(86\%) & $\mathbf{0.997}$~(81\%) \\
IMDB & $0.382$ & $0.553$ & $0.976$~(80\%) & $\mathbf{0.990}$~(90\%) & $0.983$~(84\%) \\
MMLU & $0.817$ & $0.863$ & $0.981$~(82\%) & $0.993$~(87\%) & $\mathbf{0.996}$~(84\%) \\
NYT & $0.209$ & $0.278$ & $0.990$~(40\%) & $\mathbf{0.990}$~(44\%) & $0.967$~(20\%) \\
SST-2 & $0.819$ & $0.886$ & $0.972$~(37\%) & $0.980$~(41\%) & $\mathbf{0.996}$~(56\%) \\
\bottomrule
\end{tabular}}
\end{minipage}

\vspace{1.5em}

\begin{minipage}[t]{0.48\textwidth}
\centering
\caption{$F_1$ at fixed delegation budgets. Each cell shows the best $F_1$ achievable with delegation rate $d \leq$ budget. Bold marks the best algorithm per column and dataset.}
\label{tab:fixed-budget}
\resizebox{\linewidth}{!}{%
\begin{tabular}{lcccccc}
\toprule
 & \multicolumn{3}{c}{$d \leq 20\%$} & \multicolumn{3}{c}{$d \leq 30\%$} \\
\cmidrule(lr){2-4} \cmidrule(lr){5-7}
Dataset & \supgsp & \supgit & \gamcal & \supgsp & \supgit & \gamcal \\
\midrule
ArXiv & $0.784$ & $0.798$ & $\mathbf{0.851}$ & $0.867$ & $0.873$ & $\mathbf{0.890}$ \\
BoolQ & $0.880$ & $0.893$ & $\mathbf{0.931}$ & $0.900$ & $0.922$ & $\mathbf{0.951}$ \\
IMDB & --- & --- & $\mathbf{0.685}$ & $0.770$ & $0.764$ & $\mathbf{0.771}$ \\
MMLU & $0.857$ & $0.863$ & $\mathbf{0.889}$ & $0.857$ & $0.890$ & $\mathbf{0.923}$ \\
NYT & $0.953$ & $0.946$ & $\mathbf{0.967}$ & $\mathbf{0.977}$ & $0.974$ & $0.967$ \\
SST-2 & $0.934$ & $0.923$ & $\mathbf{0.954}$ & $0.954$ & $0.964$ & $\mathbf{0.977}$ \\
\bottomrule
\end{tabular}}
\end{minipage}%
\hfill
\begin{minipage}[t]{0.48\textwidth}
\centering
\caption{Minimum delegation rate to achieve target $F_1$. Lower values indicate more cost-efficient algorithms. Bold marks the best algorithm per column.}
\label{tab:min-delegation}
\resizebox{\linewidth}{!}{%
\begin{tabular}{lcccccc}
\toprule
 & \multicolumn{3}{c}{$F_1 \geq 0.9$} & \multicolumn{3}{c}{$F_1 \geq 0.95$} \\
\cmidrule(lr){2-4} \cmidrule(lr){5-7}
Dataset & \supgsp & \supgit & \gamcal & \supgsp & \supgit & \gamcal \\
\midrule
ArXiv & $42.5\%$ & $38.9\%$ & $\mathbf{34.5\%}$ & $\mathbf{55.6\%}$ & $61.9\%$ & $61.6\%$ \\
BoolQ & $24.0\%$ & $30.2\%$ & $\mathbf{10.1\%}$ & $69.7\%$ & $58.2\%$ & $\mathbf{29.3\%}$ \\
IMDB & $\mathbf{52.6\%}$ & $52.8\%$ & $54.4\%$ & $69.9\%$ & $69.1\%$ & $\mathbf{68.4\%}$ \\
MMLU & $40.8\%$ & $41.8\%$ & $\mathbf{28.6\%}$ & $67.4\%$ & $62.9\%$ & $\mathbf{43.9\%}$ \\
NYT & $16.1\%$ & $16.0\%$ & $\mathbf{9.4\%}$ & $21.9\%$ & $22.4\%$ & $\mathbf{17.0\%}$ \\
SST-2 & $16.1\%$ & $18.3\%$ & $\mathbf{8.9\%}$ & $28.6\%$ & $31.6\%$ & $\mathbf{21.2\%}$ \\
\bottomrule
\end{tabular}}
\end{minipage}
\end{table*}

We evaluate both algorithms against four baselines on six datasets spanning classification, filtering, and join operators. The experiments address three questions: Does learned calibration improve cost-efficiency over statistical threshold estimation (Section~\ref{sec:exp-pareto})? Does \gamcal's $\alpha$ parameter provide predictable control across datasets (Section~\ref{sec:exp-alpha})? Does \supgit reliably satisfy user-specified precision-recall targets (Section~\ref{sec:exp-reliability})?

\subsection{Experimental Setup}\label{sec:exp-setup}

\paragraph{Platform.}
All experiments run on Snowflake's Cortex AISQL~\cite{liskowski2025aisql}, a production SQL engine for semantic operators. The proxy model is Llama~3.1-8B and the oracle model is Llama~3.3-70B, deployed as Cortex LLM inference endpoints. Data is processed in the streaming execution model of Section~\ref{sec:streaming} with a batch size of $|B_t| = 4{,}096$ rows and a single worker ($W = 1$). Within each batch, oracle samples are acquired in sub-batches of $128$ records, allowing both algorithms to refine thresholds multiple times per batch. Appendix~\ref{app:dop} confirms that both algorithms are robust to parallelism ($W$ up to 16).

\paragraph{Datasets.}
We select six datasets spanning different domains, task types, and proxy difficulty levels (Table~\ref{tab:datasets}). MMLU~\cite{hendrycks2020measuring} (a multiple-choice QA benchmark reduced to a binary predicate: \emph{is the selected answer correct?}), BoolQ~\cite{clark2019boolq}, and SST-2~\cite{socher2013recursive} cover knowledge QA, reading comprehension, and sentiment analysis with moderate proxy quality ($F_1 > 0.8$). In this regime, the proxy alone provides reasonable accuracy. IMDB~\cite{maas2011learning} and ArXiv~\cite{cohan2018discourse} present harder calibration challenges: on IMDB, the proxy predicts nearly every review as positive, achieving high recall but low precision ($\text{ECE} = 0.452$, the highest in the suite). ArXiv combines a highly imbalanced predicate with comparable proxy accuracy. NYT~\cite{sandhaus2008nyt} is the largest benchmark in the suite: 250K candidate pairs from an \texttt{AI\_JOIN} over article titles and excerpts, with the lowest positive rate and weakest proxy.

The expected calibration error (ECE) reported in Table~\ref{tab:datasets} quantifies the gap between proxy scores and true probabilities. High ECE motivates the learned calibration approach of \gamcal.

\paragraph{Algorithms.}
We evaluate \textbf{\supgit} (Section~\ref{sec:supgit}) and \textbf{\gamcal} (Section~\ref{sec:gamcal}) against four baselines. \textbf{\sn}~\cite{kang2020approximate} is the original recall-only cascade, optimizing a single threshold with no precision control. \textbf{\supgsp} adds joint precision-recall targeting and uncertain-region delegation to \sn but estimates thresholds from a single oracle sample per batch, without the iterative refinement loop of Algorithm~\ref{alg:supgit}. \supgsp corresponds to the cascade algorithm implemented in LOTUS~\cite{patel2025semantic}. Comparing \supgsp to \supgit thus isolates the contribution of iterative refinement. Two reference baselines anchor the cost extremes: \textbf{Proxy-only} classifies all records using the proxy model alone ($d = 0$), and \textbf{Oracle-only} delegates every record to the oracle ($d = 1.0$).

We use \sn and \supgsp as per-partition baselines rather than the more recent BARGAIN~\cite{zeighami2025bargain}. BARGAIN sharpens \sn's threshold estimates through multi-round adaptive sampling, which is harder to apply within each partition's sample budget than \sn's single pass. Section~\ref{sec:related} positions these estimator improvements relative to our streaming contributions.

\paragraph{Protocol.}
Each configuration is evaluated across 10 random seeds. To trace cost-quality tradeoff curves, we sweep each algorithm's native control parameter: \supgit and \supgsp sweep a shared target $t_P = t_R \in [0.55, 0.95]$; \sn sweeps $t_R$ over the same range; and \gamcal sweeps $\alpha \in [0.10, 0.80]$ with no budget cap ($\rho = 1.0$), so that $\alpha$ alone controls the effective delegation rate. The SUPG variants use budget fraction $\rho = 0.1$ and failure probability $\delta = 0.2$; \gamcal uses $\beta = 1$ ($F_1$ score) in the cost-quality objective. All quality metrics ($F_1$, precision, recall) are measured against oracle labels, following the convention established in Section~\ref{sec:problem}. We also report delegation rate $d$. Error bars show mean $\pm$ standard deviation across seeds.

\subsection{Cost-Quality Tradeoff}\label{sec:exp-pareto}

\begin{figure*}[t]
  \centering
  \includegraphics[width=\textwidth]{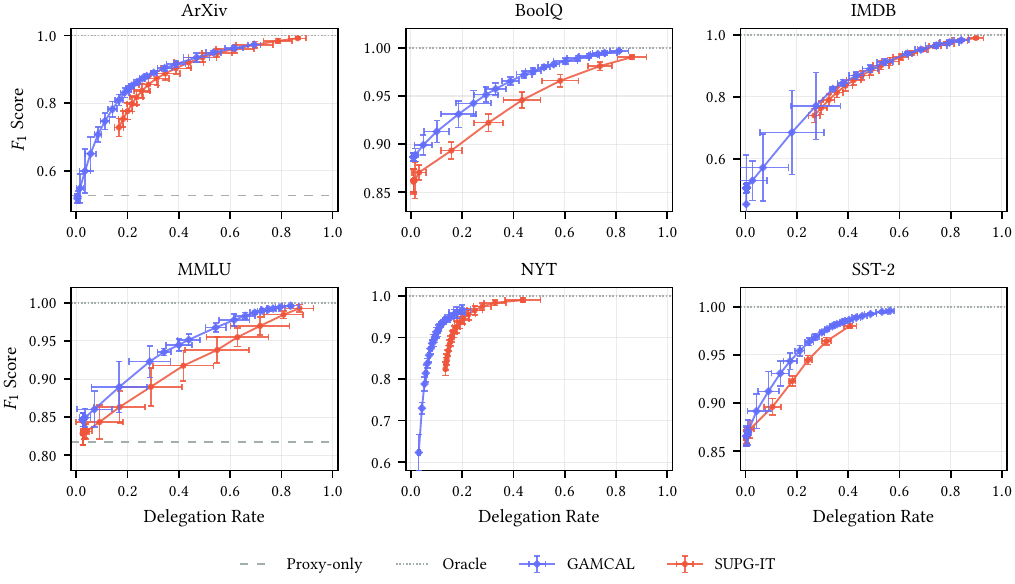}
  \caption{$F_1$ vs.\ delegation rate for \gamcal (sweeping $\alpha$) and \supgit (sweeping shared target $t_P = t_R$) across six datasets. Each point is the mean over 10 seeds (error bars: one standard deviation). Dashed horizontal lines mark the proxy-only and oracle baselines. \gamcal's frontier lies above or overlaps \supgit's on every dataset.}
  \label{fig:pareto}
\end{figure*}

Figure~\ref{fig:pareto} traces the cost-quality Pareto frontier for \gamcal and \supgit across all six datasets. At low-to-moderate delegation rates, \gamcal's frontier lies above or overlaps \supgit's on every dataset. Table~\ref{tab:best-f1} summarizes the best operating point per algorithm. \supgit reaches $F_1 \geq 0.99$ on five of the six datasets and \gamcal on three, but \gamcal peaks at lower delegation on BoolQ, MMLU, and SST-2, while \supgit reaches a higher ceiling on ArXiv, IMDB, and NYT at the cost of 6--24 percentage points more delegation.

Learned calibration confers a clear advantage at low delegation budgets. Table~\ref{tab:fixed-budget} reports the best $F_1$ achievable within delegation caps of $20\%$ and $30\%$. At $d \leq 20\%$, \gamcal leads on all six datasets, outperforming both \supgit and \supgsp (the LOTUS cascade). The widest gap is on ArXiv: $0.851$ vs.\ $0.784$ for \supgsp and $0.798$ for \supgit. On IMDB, the SUPG variants cannot operate within this budget at all because their minimum delegation includes both the $\rho = 10\%$ sampling budget and mandatory uncertain-region delegation. Raising the cap to $d \leq 30\%$ narrows the gap: \gamcal still leads on five of six datasets, while \supgsp edges ahead on NYT ($0.977$ vs.\ $0.967$) where \gamcal's delegation ceiling limits further gains.

Table~\ref{tab:min-delegation} examines the same tradeoff from the opposite direction. \gamcal requires less delegation than \supgit to reach a target $F_1$ on all six datasets for $F_1 \geq 0.95$ and on five for $F_1 \geq 0.90$. The advantage over \supgsp is larger still: on BoolQ, reaching $F_1 \geq 0.95$ costs $d = 29.3\%$ with \gamcal versus $69.7\%$ with \supgsp, a $58\%$ reduction in oracle calls. On MMLU, the savings are $35\%$ ($43.9\%$ vs.\ $67.4\%$). The GAM calibration model generalizes across the proxy score distribution: labels at one score level inform predictions at others. The sample-based bounds used by \supgit and \supgsp extract less information from the same number of oracle labels.

\begin{figure*}[t]
    \centering
    \includegraphics[width=0.83\textwidth]{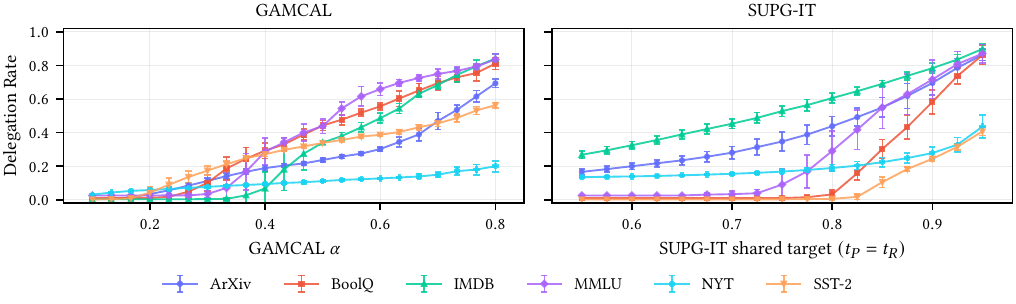}
    \caption{Delegation rate as a function of each algorithm's native control parameter across all six datasets. Left: \gamcal sweeps $\alpha$, where higher $\alpha$ prioritizes classification quality over oracle cost. Right: \supgit sweeps the shared target $t_P = t_R$. Each point is the mean over 10 seeds (error bars: one standard deviation). The $\alpha$ parameter produces smooth monotonic curves across datasets. A target specifies desired quality, and the delegation required to achieve it varies with dataset difficulty.}
    \label{fig:alpha-guidance}
  \end{figure*}

Among individual datasets, NYT best illustrates \gamcal's cost advantage. It crosses $F_1 \geq 0.90$ at under $10\%$ delegation, the lowest threshold in the suite. \supgit reaches a higher ceiling ($F_1 = 0.990$ at $d = 44\%$) because the GAM classifies most NYT records confidently in this extreme-imbalance setting and caps delegation at roughly $20\%$ regardless of $\alpha$. IMDB presents the hardest calibration challenge. The proxy scores $F_1 = 0.382$ alone, predicting nearly every review as positive. Both algorithms nevertheless recover to near-oracle quality ($F_1 > 0.98$) at high delegation.

Both algorithms substantially outperform \sn, which is limited to $F_1 \leq 0.90$ and offers no precision control. The limitation is inherent rather than budget-related. \sn optimizes a single threshold for recall only. Increasing the sampling budget $\rho$ therefore tightens the threshold estimate but cannot address the absence of precision control that causes low $F_1$ on datasets with high false-positive rates (e.g., NYT).

Within the SUPG family, adding joint precision-recall targeting and uncertain-region delegation (the step from \sn to \supgsp) produces the largest quality gain. \supgsp's peak $F_1$ ranges from $0.972$ (SST-2) to $0.990$ (NYT), far above \sn's ceiling of $0.278$--$0.899$. Iterative refinement (the step from \supgsp to \supgit) adds 1--2 $F_1$ points on all datasets except NYT, where the clean binary join signal makes single-pass estimation already optimal. Appendix Figure~\ref{fig:pareto-grid} provides the full comparison across all four cascade algorithms and all four metrics.

In summary, the two proposed algorithms occupy complementary niches. \gamcal achieves equal or higher $F_1$ per oracle call at cost-sensitive operating points, without requiring users to specify quality targets. Its calibration model is particularly effective when few oracle labels are available. \supgit reaches a higher quality ceiling on datasets with challenging calibration (ArXiv, IMDB, NYT) and provides explicit probabilistic guarantees on precision and recall. Sections~\ref{sec:exp-alpha} and~\ref{sec:exp-reliability} examine these distinct strengths, evaluating \gamcal's parameter predictability and \supgit's target reliability.

\subsection{Parameter Predictability}\label{sec:exp-alpha}

Section~\ref{sec:gamcal-motivation} argued that target-based control requires \emph{a priori} knowledge of dataset difficulty, since the same target can yield unpredictable delegation rates across workloads. Figure~\ref{fig:alpha-guidance} tests this claim by plotting each algorithm's delegation rate against its native control parameter.

\gamcal's $\alpha$-to-delegation mapping (left panel) is smooth and monotonic for every dataset, so practitioners can treat $\alpha$ as a predictable cost dial. The spread across datasets at a given $\alpha$ reflects automatic adaptation to proxy quality: at $\alpha = 0.5$, IMDB receives $34\%$ delegation while NYT receives only $11\%$. Even at $\alpha = 0.8$, NYT's delegation plateaus at roughly $20\%$. The GAM classifies most records confidently in this extreme-imbalance join setting, and further delegation cannot improve quality.

\supgit's target-to-delegation mapping (right panel) reflects qualitatively different parameter semantics. A target specifies \emph{desired quality}, and the delegation required to achieve it depends on the proxy's intrinsic accuracy: at $t_P = t_R = 0.80$, delegation ranges from $0.6\%$ on SST-2 to $60.6\%$ on IMDB. On easier datasets, the proxy alone satisfies low targets, so the flat regions in the right panel represent configurations where adjusting the target has no effect on cost. On SST-2, delegation stays below $1\%$ for targets up to $0.75$ before rising to $24\%$ at $t_P = t_R = 0.90$. Delegation rises sharply only when the target exceeds what the proxy can achieve alone.

\subsection{Target Reliability}\label{sec:exp-reliability}

\begin{table}[t]
    \centering
    \caption{\supgit target reliability across 289 $(t_P, t_R)$ pairs per dataset, 10 seeds each (2{,}890 runs per dataset). The $t_P{=}t_R$ column reports the 17 symmetric pairs only (170 runs).}
    \label{tab:reliability}
    \begin{tabular}{lrrrr}
      \toprule
      & \multicolumn{2}{c}{Satisfaction (\%)} & & \\
      \cmidrule(lr){2-3}
      Dataset & All & $t_P{=}t_R$ & Failures & Delegation (\%) \\
      \midrule
      ArXiv & 100.0 & 100.0 &   0 & 41.4 \\
      IMDB  & 100.0 & 100.0 &   0 & 54.6 \\
      MMLU  &  99.4 & 100.0 &  16 & 23.6 \\
      BoolQ &  98.7 &  99.4 &  38 & 14.4 \\
      SST-2 &  89.4 &  99.4 & 305 &  4.7 \\
      \bottomrule
    \end{tabular}
\end{table}

Section~\ref{sec:exp-alpha} examined \gamcal's parameter predictability. Here we evaluate whether \supgit reliably delivers the joint precision-recall targets the user specifies. A $17 \times 17$ grid of target pairs $(t_P, t_R)$ from $0.55$ to $0.95$ in steps of $0.025$ is swept across five datasets with 10 random seeds each, for a total of 14{,}450 runs. NYT is excluded due to the computational cost of the full grid on 250K rows. Table~\ref{tab:reliability} summarizes the results. ArXiv and IMDB achieve perfect joint satisfaction across all 289 configurations. These are also the highest-delegation datasets.

BoolQ and MMLU maintain per-run satisfaction of at least 98.7\%, with failures confined to low-delegation configurations. SST-2 shows the most target misses (89.4\% satisfaction), concentrated in regions where the proxy's intrinsic quality nearly satisfies the target with minimal oracle involvement. Among failing configurations, mean delegation is only 1.4\%. The mechanism is consistent: low delegation yields insufficient oracle data for tight statistical bounds, and seed-to-seed variation produces occasional misses. Every failure in the experiment violates exactly one metric (precision or recall), never both. Appendix~\ref{app:reliability-heatmaps} visualizes the spatial pattern of these failures.

Across the full grid, every dataset meets both targets jointly in at least 89.4\% of runs, so each target holds individually at least that often, above the $1 - \delta = 80\%$ reliability that \supgit targets at $\delta = 0.2$. The symmetric targets used in Sections~\ref{sec:exp-pareto}--\ref{sec:exp-alpha} achieve near-perfect satisfaction on all five datasets (Table~\ref{tab:reliability}, $t_P{=}t_R$ column). Reliability degrades primarily in asymmetric configurations at low delegation levels, precisely where the cascade adds the least value.

\section{Conclusion}\label{sec:conclusion}

We formalized the model cascade problem for semantic SQL in a streaming execution model with independent parallel workers and presented two complementary algorithms. \supgit extends SUPG to streaming execution with iterative threshold refinement and joint precision-recall guarantees. Each worker processes its partition independently, requiring no global synchronization. \gamcal replaces user-specified targets with a learned calibration model that directly optimizes a cost-quality tradeoff and adapts automatically to dataset difficulty. Experiments on six datasets spanning different domains, task types, and proxy quality levels confirm that both algorithms outperform existing baselines, including the SUPG cascade in LOTUS~\cite{patel2025semantic}. \gamcal achieves higher $F_1$ per oracle call at cost-sensitive operating points, requiring up to $58\%$ fewer oracle calls than LOTUS to reach $F_1 \geq 0.95$, while \supgit reaches a higher quality ceiling with a mean peak $F_1$ of $0.989$ and provides probabilistic guarantees on precision and recall. For practitioners, \gamcal is the default choice for cost-sensitive workloads where no specific quality target is required, while \supgit is preferred when probabilistic guarantees on precision and recall are needed.

The approach has two limitations. Our formulation treats oracle labels as ground truth, an assumption that holds when the oracle is considerably more accurate than the proxy but may degrade under noisy or adversarial oracle conditions. The current framework handles binary predicates only, whereas production semantic SQL engines also support multi-class operators that require extending the two-threshold decision framework.

Generalizing the cascade to multi-class settings is a natural next step. The union-bound composition of per-worker guarantees (Section~\ref{sec:streaming}) is conservative. Tighter global bounds that exploit partition structure could reduce the per-worker failure probability budget. Cross-partition coordination mechanisms that preserve global quality guarantees without inter-worker communication and richer calibration models that adapt to non-stationary data distributions are further promising directions.

\section*{Acknowledgments}
We are deeply grateful to Filip Grali\'nski and Dimitris Tsirogiannis, whose careful reading and thoughtful suggestions shaped this paper for the better. We thank them for their time and their insight.

\clearpage
\bibliographystyle{abbrv}
\bibliography{references}

\clearpage
\appendix

\section{Calibration Model Variants}\label{app:calibration-variants}

Section~\ref{sec:gamcal-calibration} presents the GAM calibration framework and derives the stochastic calibrated score (Eq.~\ref{eq:calibrated-score}). The framework admits three implementations, each offering different tradeoffs between monotonicity guarantees, computational cost, and uncertainty robustness.

\subsection{Constrained GAM with Analytical Confidence Intervals}\label{app:constrained-gam}

All experiments in Section~\ref{sec:experiments} use this variant as the default calibration model. The implementation represents $f$ as a cubic spline and fits the penalized log-likelihood (Eq.~\ref{eq:gam-likelihood}) subject to monotonicity constraints $g'(s) \geq 0$, enforced as linear inequality constraints on the spline coefficients during iteratively reweighted least squares (IRLS) optimization. We use the \texttt{pyGAM} library~\cite{serven2018pygam}, which implements this procedure following Wood~\cite{wood2017generalized}.

The fitted model provides a posterior approximation via the Bayesian interpretation of penalized splines~\cite{wood2017generalized}. Concretely, the roughness penalty $\lambda \int (f'')^2\, ds$ corresponds to a Gaussian prior on the spline coefficients, and the penalized likelihood yields an approximate posterior. From this posterior, the mean $\hat{f}(s)$ and standard error $\text{se}(s)$ are extracted, and the stochastic calibrated score follows Eq.~\ref{eq:calibrated-score}:
\begin{equation}
    \tilde{g}(s, q) = \text{logit}^{-1}\!\left(\hat{f}(s) + \Phi^{-1}(q) \cdot \text{se}(s)\right)
\end{equation}

\subsection{Regularized Logistic Regression with Spline Basis}\label{app:regularized-lr}

An alternative variant replaces the constrained GAM with logistic regression over an explicit spline basis expansion, offering tighter control over the optimization and a closed-form Laplace approximation for uncertainty.

Raw proxy scores are first transformed to log-odds space $\ell = \text{logit}(A(x))$, clipped to $[\ell_{\min}, \ell_{\max}]$ for numerical stability. A degree-3 B-spline basis with $p$ uniformly spaced knots is constructed via \texttt{scikit-learn}'s \texttt{SplineTransformer}, yielding $d$ basis functions $\phi_1(\ell), \ldots, \phi_d(\ell)$. The calibration function is then:
\begin{equation}
    f(s) = \sum_{j=1}^{d} \theta_j \, \phi_j(\text{logit}(s))
\end{equation}
and the coefficients $\theta$ are fit by minimizing the regularized negative log-likelihood:
\begin{equation}
    \min_\theta \; -\sum_{i=1}^{|\SSe|} \left[ y_i \log g(A(x_i)) + (1-y_i) \log(1 - g(A(x_i))) \right] + \frac{\lambda}{2} \|\theta - \mu_\theta\|^2
    \label{eq:regularized-lr}
\end{equation}
using L-BFGS-B optimization with analytically computed gradients.

The prior mean $\mu_\theta$ encodes an inductive bias toward the identity mapping in log-odds space (no calibration adjustment): the coefficients are set to linearly spaced values $\mu_j = \ell_{\min} + (\ell_{\max} - \ell_{\min}) \cdot j / d$. When oracle samples are scarce and $\lambda$ is large, the regularization pulls $\theta$ toward $\mu_\theta$, effectively recovering Platt scaling as a default. As more samples accumulate and $\lambda$'s relative influence diminishes, the model departs from this prior to capture nonlinear calibration structure. The regularized variant thus interpolates between Platt scaling (few samples) and a flexible spline model (many samples).

Confidence intervals are obtained via the Laplace approximation. At the fitted optimum $\hat{\theta}$, the Hessian of the regularized loss is:
\begin{equation}
    H = X^\top \text{diag}(\hat{h} \circ (1 - \hat{h})) \, X + \lambda I
\end{equation}
where $X$ is the spline design matrix and $\hat{h} = \sigma(X\hat{\theta})$ are the fitted probabilities. The approximate posterior covariance is $\Sigma = H^{-1}$, and the standard error of $f(s)$ at a new point is $\text{se}(s) = \sqrt{\phi(s)^\top \Sigma \, \phi(s)}$ where $\phi(s)$ is the spline basis vector. The stochastic calibrated score is computed as in Eq.~\ref{eq:calibrated-score}.

Compared to the constrained GAM, this variant does not enforce monotonicity as a hard constraint. Instead, the monotonic prior $\mu_\theta$ encourages (but does not guarantee) monotonicity. In practice, the smoothness of the spline basis and the regularization toward $\mu_\theta$ produce approximately monotone calibration curves for reasonable values of $\lambda$.

\subsection{Bootstrap Ensemble}\label{app:bootstrap}

The bootstrap variant quantifies calibration uncertainty empirically rather than analytically. A primary GAM $g$ is trained on the full oracle sample $\SSe$ as in Appendix~\ref{app:constrained-gam}. Additionally, $B$ bootstrap GAMs $g_1, \ldots, g_B$ are trained on resamples $\SSe_1^*, \ldots, \SSe_B^*$ drawn with replacement from $\SSe$ (we use $B = 100$).

Uncertainty is estimated from the ensemble spread in logit space. Let $\bar{\ell}(s)$ denote the mean bootstrap logit and $\Delta_b(s)$ the deviation of the $b$-th ensemble member:
\begin{align}
    \bar{\ell}(s) &= \frac{1}{B} \sum_{b=1}^{B} \text{logit}(g_b(s)) \\
    \Delta_b(s) &= \text{logit}(g_b(s)) - \bar{\ell}(s)
\end{align}
The centered deviations $\{\Delta_b(s)\}_{b=1}^B$ form an empirical distribution of calibration uncertainty at score level $s$. The stochastic calibrated score is:
\begin{equation}
    \tilde{g}(s, q) = \text{logit}^{-1}\!\left( \text{logit}(\mu(s)) + Q_q\!\left(\{\Delta_b(s)\}_{b=1}^B\right) \right)
    \label{eq:bootstrap-calibrated}
\end{equation}
where $\mu(s) = g(s)$ is the primary model's prediction and $Q_q$ denotes the $q$-th quantile of the empirical distribution.

Unlike the analytical CIs and Laplace approximation, the bootstrap makes no distributional assumptions about calibration uncertainty. However, it is computationally more expensive: each retraining event requires fitting $B + 1$ GAMs, and each calibration query requires $B + 1$ forward passes, roughly two orders of magnitude more than the primary variant. The bootstrap is most appropriate when the Gaussian approximation may be inadequate, e.g., with highly skewed class distributions or very few oracle samples, or when the calibration curve has sharp features that the analytical CIs may underestimate.

\begin{figure*}[t]
    \centering
    \includegraphics[width=\textwidth]{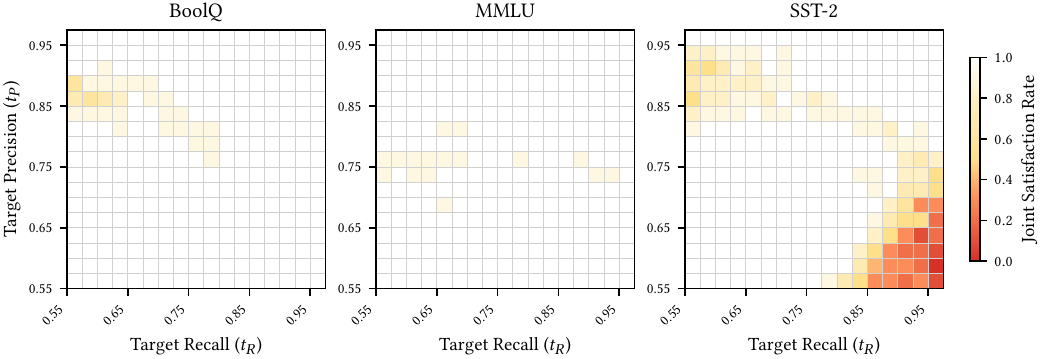}
    \caption{Joint target satisfaction rate for \supgit across a $17 \times 17$ grid of $(t_P, t_R)$ targets on BoolQ, MMLU, and SST-2 (10 seeds per configuration). White indicates 100\% satisfaction. Yellow and red indicate partial or zero satisfaction. ArXiv and IMDB (not shown) achieve 100\% on all 289 configurations.}
    \label{fig:reliability-heatmaps}
\end{figure*}

All experiments in this paper use the constrained GAM (Appendix~\ref{app:constrained-gam}) as the default because it combines monotonicity guarantees, analytical uncertainty estimates, and minimal computational overhead. The regularized variant (Appendix~\ref{app:regularized-lr}) is preferable when oracle samples are scarce: its Platt scaling prior provides a principled fallback that prevents overfitting before sufficient data accumulates. The bootstrap (Appendix~\ref{app:bootstrap}) trades a roughly $100\times$ increase in computation for distribution-free uncertainty estimates, making it appropriate when the Gaussian posterior approximation may be inadequate. The \gamcal framework is agnostic to the calibration backend: any implementation that provides calibrated probabilities with pointwise uncertainty can be substituted without modifying the cascade logic.

\section{Extended Pareto Analysis}\label{app:pareto-grid}

\begin{figure*}[t]
    \centering
    \includegraphics[width=\textwidth,height=0.9\textheight,keepaspectratio]{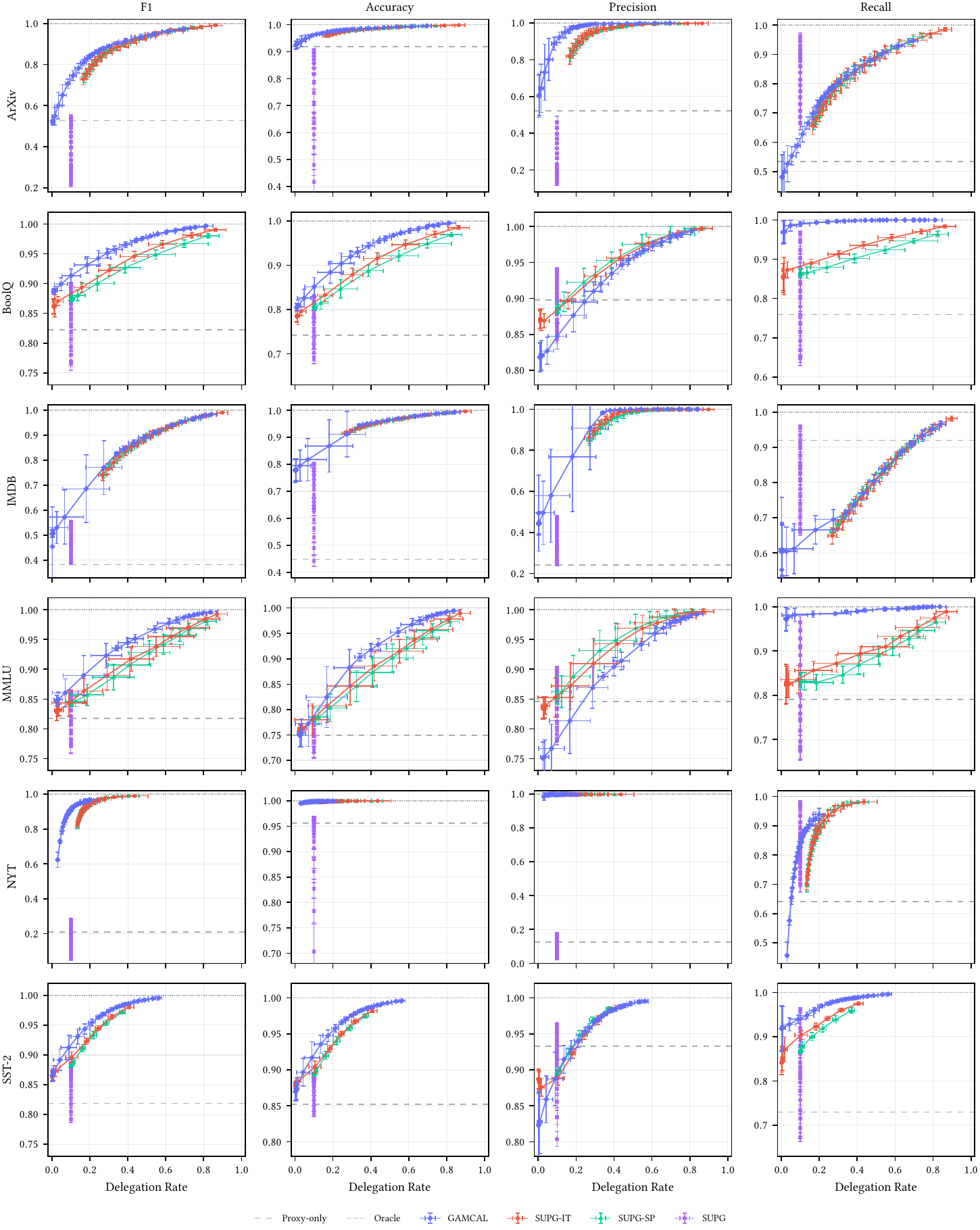}
    \caption{Complete Pareto grid: all four cascade algorithms (\gamcal, \supgit, \supgsp, \sn) across six datasets and four metrics ($F_1$, accuracy, precision, recall). Dashed horizontal lines mark the proxy-only and oracle baselines.}
    \label{fig:pareto-grid}
  \end{figure*}

Figure~\ref{fig:pareto-grid} extends the main Pareto analysis (Figure~\ref{fig:pareto}) from $F_1$ to all four quality metrics and from two algorithms to all four.

\sn's curves are vertical lines at $d = \rho = 0.1$ because the algorithm samples a fixed fraction $\rho$ of records for oracle labeling, regardless of the recall target. The oracle labels both estimate the recall threshold $\tau$ and provide final classifications for the sampled records. The proxy classifies the remainder using $\tau$. Because $\tau$ depends on $t_R$ but the sample size does not, sweeping $t_R$ changes quality metrics but not cost. By contrast, \supgsp and \supgit add a second stage that delegates all uncertain-region records to the oracle, creating a variable cost that grows with the target.

The precision and recall columns together expose the asymmetry that motivates joint targeting. \sn controls recall only: as $t_R$ increases, the threshold drops to accept more records, and precision degrades in proportion. On NYT (positive rate $0.9\%$), precision falls from $0.17$ to $0.03$ across the sweep, producing $F_1$ below $0.06$ at the highest recall targets. ArXiv and IMDB show the same pattern, with precision dropping below $0.15$ and $0.25$. Adding the upper threshold eliminates this failure mode: both \supgsp and \supgit maintain precision above $0.99$ on NYT while achieving the same recall range. The recall column confirms the complementary view: all SUPG variants achieve high recall at the strongest targets because recall is directly optimized through the lower threshold.

The dashed baseline markers reveal how each algorithm positions relative to the proxy-only and oracle bounds. At high recall targets, \sn drops below the proxy-only accuracy baseline on most datasets. ArXiv is the most extreme case: all operating points fall below the proxy line. Lowering the threshold to capture more positive records admits enough false positives to degrade overall accuracy below what the unmodified proxy achieves. The three algorithms with precision control (\gamcal, \supgit, \supgsp) avoid this degradation and approach oracle quality at high delegation, reaching $F_1$ within a few percent of perfect on every dataset.

\section{Target Satisfaction Patterns}\label{app:reliability-heatmaps}

Figure~\ref{fig:reliability-heatmaps} visualizes the spatial distribution of target failures from the experiment described in Section~\ref{sec:exp-reliability}. The failure patterns differ across datasets but share a common cause. On BoolQ, failures cluster at high precision targets ($t_P \geq 0.75$) with low recall targets, where the proxy's intrinsic precision nearly meets the target and the cascade delegates few records to the oracle. The pattern on MMLU is milder: all 16 failures reach $90\%$ satisfaction, again concentrated in low-delegation regions. About a third of configurations on SST-2 fall short of full satisfaction. The worst cases occur at high recall with low precision targets (e.g., $t_P = 0.575$, $t_R = 0.95$), where delegation drops below $1\%$ and precision is the metric that misses in every case. Across all three datasets, failures concentrate where the cascade is barely active: mean delegation in failing configurations is $1$--$8\%$. The proxy alone nearly satisfies the targets in these regions, and the few oracle samples collected are insufficient for tight statistical bounds.

\section{Robustness to Parallelism}\label{app:dop}

Both algorithms are designed for independent per-worker execution. Increasing the number of workers $W$ (the degree of parallelism) splits the data into smaller partitions. Each worker therefore runs fewer iterative refinement steps and may face partition-level class imbalance. To quantify this effect, we evaluated \gamcal and \supgit at $W \in \{1, 2, 4, 8, 16\}$ on IMDB, MMLU, and SST-2 with batch size $4096$ and 10 seeds per configuration.

Quality is robust to parallelism. The mean best $F_1$ across datasets varies by less than $0.001$ for \gamcal and less than $0.004$ for \supgit as $W$ increases from 1 to 16. \gamcal is the more stable of the two, with a maximum per-dataset shift of $0.002$ $F_1$. \supgit shows a larger shift on SST-2 ($+0.012$ $F_1$ at $W{=}16$), where fewer batches per worker leave thresholds less converged, widening the uncertain region and raising delegation from $41\%$ to $56\%$. The additional oracle labels improve quality at higher cost. Seed-level variability does not increase with $W$: for both algorithms, the standard deviation of $F_1$ across seeds remains stable or decreases as partitions shrink.

\end{document}

%% file: tikz-streaming-model.tex
\begin{figure}[t]
\centering
\begin{tikzpicture}[
    worker/.style={rectangle, draw, minimum width=2.2cm, minimum height=1.6cm, font=\small, align=center},
    arrow/.style={->, >=stealth, thick},
    batch/.style={rectangle, draw, dashed, minimum width=1.4cm, minimum height=0.5cm, font=\scriptsize},
    >=stealth
]

\node[worker] (w1) at (0, 0) {Worker $W_1$\\[-2pt]\scriptsize\textit{local state}\\[-2pt]\scriptsize$\tau_{\text{low}}^{(1)}, \tau_{\text{high}}^{(1)}$};
\node[worker] (w2) at (3.2, 0) {Worker $W_2$\\[-2pt]\scriptsize\textit{local state}\\[-2pt]\scriptsize$\tau_{\text{low}}^{(2)}, \tau_{\text{high}}^{(2)}$};
\node[worker] (w3) at (6.4, 0) {Worker $W_3$\\[-2pt]\scriptsize\textit{local state}\\[-2pt]\scriptsize$\tau_{\text{low}}^{(3)}, \tau_{\text{high}}^{(3)}$};

\node[batch] (b1) at (0, 1.8) {$B_1^{(1)}, B_2^{(1)}, \ldots$};
\node[batch] (b2) at (3.2, 1.8) {$B_1^{(2)}, B_2^{(2)}, \ldots$};
\node[batch] (b3) at (6.4, 1.8) {$B_1^{(3)}, B_2^{(3)}, \ldots$};

\draw[arrow] (b1) -- (w1);
\draw[arrow] (b2) -- (w2);
\draw[arrow] (b3) -- (w3);

\node[font=\scriptsize] (out1) at (0, -1.6) {predictions};
\node[font=\scriptsize] (out2) at (3.2, -1.6) {predictions};
\node[font=\scriptsize] (out3) at (6.4, -1.6) {predictions};
\draw[arrow] (w1) -- (out1);
\draw[arrow] (w2) -- (out2);
\draw[arrow] (w3) -- (out3);

\end{tikzpicture}
\caption{Streaming execution model. Each worker processes its data partition independently, maintaining local threshold estimates and updating them based on its own oracle observations. Workers do not share samples or synchronize.}
\label{fig:streaming-model}
\end{figure}